\documentclass[11pt]{elsarticle}
\usepackage[english]{babel}
\usepackage{booktabs}
\usepackage{adjustbox}
\usepackage{blindtext}
\usepackage{pifont}
\usepackage{float}
\usepackage{hyperref}
\usepackage{rotfloat}
\usepackage{natbib}
\usepackage{bbm}
\usepackage{bm}
\usepackage{amsmath}
\usepackage{amssymb}
\usepackage{subcaption}
\usepackage{caption}
\usepackage{float}
\begin{document}
\begin{frontmatter}
\title{\textbf{Binning as a Regularization Factor: \\
A comparative study of Richardson-Lucy Deconvolution and Data Unfolding with Mean Integrated Squared Error Optimization}}
\vspace {-0.1cm}
\author{Nikolay D.\ Gagunashvili\corref{author}}
\address{University of Iceland, S\ae mundargata 2, 101 Reykjavik, Iceland}
\begin{abstract}
\vspace {-0.1cm}
Experimental data in particle and nuclear physics, particle astrophysics, and radiation protection dosimetry are collected using experimental facilities consisting of complex systems of sensors, electronics, and software.
Measured spectra or cross sections are considered probability density functions (PDFs) that deviate from the true PDFs due to detector resolution, bias, and efficiency effects.

Reliable estimates of the true PDF, obtained through unfolding, are essential for testing theoretical models, comparing results from different experiments, and combining results from various research studies.
Both external and internal methods can be applied to assess the quality of unfolding results.
A typical example of the application of an external quality assessment method is the deconvolution of a blurred image, where the sharpness of the restored image serves as an indicator of quality. However, defining such external criteria can be challenging, particularly when a measurement has not been performed previously.

This paper discusses various internal criteria for assessing the quality of unfolding results independently of external information, as well as the factors influencing the quality of the unfolded distribution.
This work extends our previous comparative study by systematically investigating the influence of binning choice on the quality of Richardson--Lucy deconvolution and data unfolding based on mean integrated squared error optimization.
Furthermore, a new method for calculating the elements of the response matrix with minimal variances, is developed.
\end{abstract}
\end{frontmatter}
\vspace{2cm}
\section {Introduction}
A measurement is produced by an experimental setup comprising  a complex set of sensors, electronics, and software. 
The measured Probability Density Function (PDF) $f (y)$ differs from the true PDF  $\phi(x)$   due to
the resolution of an  experimental setup, the recording efficiency, and other factors.

As a typical example, noise components are added to the kinematic parameters of a particle after it passes through  the material of 
 a detector. In some cases may  not be recorded at all because of its energy loss.

The problem of estimating an unknown PDF  $\phi(x)$ which may represent  a spectrum or  a differential cross section, is one of the main subjects of analysis in particle physics, nuclear physics, particle astrophysics, and   radiation protection dosimetry.  It is common  part of data analysis and traditionally  referred to as unfolding.  Computer modeling is often employed to study the registration process.\\

As discussed  in \cite{ng, Gagunashvili_2026}, the data under analysis are represented by two sets of random variables:
\begin{enumerate}
\item The measured data sample is a collection of independent, identically distributed (IID) of $n$ random variables:
\begin{equation}
 y_1, y_2,\ldots ,y_{n} \label{meas}
\end{equation}
with the  Probability Density Function (PDF) $f (y)$.
\begin{itemize}
\item The true PDF  $\phi(x)$ is not known.
\item The measured PDF $f(y)$  is also unknown, however, it can be estimated.
\end{itemize}
\item The simulated data sample is a collection of $k$ pairs of IID random variables:
\begin{equation}
x_1^s, y_1^s ; \, x_2^s, y_2^s; \, \ldots \,x_k^s, y_k^s  \label{simul}   
\end{equation}
with  PDF's  $\phi^s(x)$ and  $f^s(y)$. 
\begin{itemize}
\item The generated PDF  $\phi^s(x)$ is an analog of the true PDF $\phi(x)$.  
In the common case,  $\phi^s(x) \ne \phi (x)$ and the PDF $\phi^s(x)$  can be estimated. 
\item The reconstructed PDF $f^s (x)$  is an analog of the measured PDF $f(x)$ and  the PDF   $f^s (x)$ can be estimated. 
 \end{itemize}
\end{enumerate}
The simulated sample serves the purpose of constructing a mathematical model for the measurement system. 

A commonly employed model for this case is the Fredholm integral equation, which describes  the relationship between the measured PDF denoted as $f(y)$ and the true PDF  $ \phi(x)$.
\begin{equation}
\int \limits_{-\infty}^{+\infty} R(x,y)A(x) \phi(x) dx=f(y),  \,\,    \label{fred}
\end {equation}
where $A(x)$ is the probability of recording of an event with a characteristic $x$
(the acceptance); $R(x,y)$, is the probability of obtaining $y$ instead of $x$ (the
experimental resolution).
In  the special  case where $R(x,y)=R(y-x)$, a formal solution can be expressed using Fourier transform as follows:
\begin{equation}
  \phi(x) = \frac{1}{A(x)} \int\limits_{-\infty}^{+\infty} e^{ipx}\frac{\widetilde{f}(p) }{ \widetilde{R}(p)}dp, \label{fur}
\end {equation}
where the sign $\sim$ denotes the Fourier transform of the corresponding functions. 

If the Fourier transform of the kernel, $\widetilde{R}(p)$, vanishes within  the high-frequency region of the spectrum,  this implies that the measured PDF 
$f(y)$ does not contain information about the high-frequency behavior of $\phi(x)$. Formally, it is not possible to determine  $\phi(x)$
from equation (\ref{fur}) in this frequency domain, just as it is impossible to define $\phi(x)$ in regions where the acceptance $A(x)$ equals  zero. 

Fredholm integral equations \ref{fred}) are therefore ill-posed. Regularization addresses the difficulties described above by restricting the admissible solution space, thereby transforming the ill-posed problem into a well-posed one.

The Fredholm equation example demonstrated the liner dependence   of the measured PDF $f(y)$ on the true PDF $\phi(x)$. 
Although this linear dependence is reasonable, in many practical cases it requires careful investigation to determine whether this dependence is truly linear or instead nonlinear.
\section {Quality assessment  of the unfolding  procedure}
 Both external and internal criteria are employed to assess the quality of the unfolding procedure. In certain instances, external criteria are available, providing a framework for evaluating the reliability of the procedure. However, in experimental physics, defining external criteria can be challenging, particularly when no prior measurements exist. In such scenarios, model-based assumptions often serve as a guide. Conversely, internal criteria for evaluating the quality of the unfolding result become essential when external references are unavailable.

The quality criteria for the data unfolding procedure presented below are integral to the evaluation process. 
Notably, these criteria not only help determine the optimal values of the parameters of unfolding algorithms but also facilitate the comparison of different algorithms.

The unfolded distribution  is an estimator of the true distribution of $\phi(x)$ and therefore differs from it.
In order to develop a method that produces an estimator  $\hat{\phi}(x)$ that is close as possible to  $\phi(x)$,  appropriate measures must be defined to quantify this discrepancy.

\subsection{Mean Integrated Square Error}
One commonly used measure of the accuracy of the estimator   $\hat{\phi}(x)$  is the Mean Integrated Square Error (MISE) \cite{silver, Tsybakov2009, Scott2015, WandJones1995}.
To define MISE,  we first introduce the Integrated Square Error (ISE):
\[
\mathrm{ISE}(\hat{\phi}(x) ) = \int \big( \hat{\phi}(x) - \phi(x)\big)^2 \, dx.
\]
Then MISE can be written as 
\[
\mathrm{ MISE}(\hat{\phi}(x) ) = \mathbf{E}(\mathrm{ISE})= \mathbf{E}\!\left[ \int \big( \hat{\phi}(x) - \phi(x)\big)^2 \, dx \right].
\]
It is also possible give an equivalent formula for MISE: 
\[
\mathrm{MISE}(\hat{\phi}(x) ) = \int \mathbf{E}\!\left[ \big( \hat{\phi}(x) - \phi(x)\big)^2 \right] \, dx.
\]
Substituting the formula 
\[
\mathbf{E}\!\left[ \big( \hat{\phi}(x) - {\phi}(x)  \big)^2 \right]
= \big( \mathrm{Bias}\,\big[\hat{\phi}(x)  \big] \big)^2+\mathrm{Var}\!\big( \hat{\phi}(x)  \big)
\]
into the above expression,  we obtain  a useful representation of MISE:
\[
\mathrm{MISE}(\hat{\phi}(x)  ) = \int\, \left[ \big( \mathrm{Bias}\,\big[ \hat{\phi}(x)  \big] \big)^2+ \mathrm{Var}\!\big( \hat{\phi}(x)  \big) \right] dx.
\]
An unfolded distribution that minimizes the Mean Integrated Square Error  achieves an optimal trade-off between bias and variance.
An algorithm with the lowest value of MISE is therefore preferable.

Below we consider internal quality criteria for  an unfolding distribution  represented through  step-function approximation. 

Let us define the positions of the  $m$ steps as $\{a_1, a_2,..., a_{m+1}\}$
and the probability density function  estimate of the unfolded distribution  as
\begin{equation}
\hat{\phi}(x)=\frac {\hat{\phi_i}}{a_{i+1}-a_{i}} \, \,\text{for} \, \, \, a_{i} \leq x <  a_{i+1}, 
\end{equation}
where $\hat{\phi_i}$,  is  an estimator  of  
\begin{equation}
 \phi_i= \int\limits_{a_i}^{a_{i+1}} \phi(x)dx. 
\end{equation}
In this case, the MISE is expressed as
\begin{equation}
 \begin{aligned}
\mathrm{MISE} = & \int \limits_{a_1}^{a_{m+1}}\mathbf{E} \,[\hat{\phi }(x)^2] dx- 2 \int \limits_{a_1}^{a_{m+1}} \phi(x)  \mathbf{E}[\hat{\phi }(x)] dx+  \int \limits_{a_1}^{a_{m+1}}\phi (x)^2dx\\
=& \sum_{i=1}^m \frac { \mathbf{E} \hat{\phi_i}^2} { a_{i+1}-a_{i}} 
- 2  \sum_{i=1}^m   \frac {\mathbf{E}\hat{\phi_i}} { a_{i+1}-a_{i}} \phi_i+ \int \limits_{a_1}^{a_{m+1}}\phi (x)^2dx. \label{msem}
\end{aligned}
\end{equation}
The last term in expression  (\ref{msem}) does not depend on the estimators  $\hat{\phi_i}$; it is a constant. Therefore, the choice of parameters and factors that minimize the MISE does not depend on this part of the equation.

Another well-known and useful representation of MISE  is
\begin{equation}
\mathrm{MISE}
=
\sum_{i=1}^m \frac{1}{a_{i+1}-a_{i}}\mathbf{E}\!\left[ (\hat \phi_i - \phi_i)^2 \right]
+
 \sum_{i=1}^m\int_{a_i}^{a_{i+1}} \left( \phi(x) - \frac{\phi_i}{a_{i+1}-a_{i}} \right)^2 \, dx
\end{equation}
or equivalently,
\[
\mathrm{MISE}
=
 \sum_{i=1}^m\frac{1}{a_{i+1}-a_i}
\left[
\big( \mathrm{Bias}\,\big[\hat{\phi_i}  \big] \big)^2
+
\operatorname{Var}(\hat \phi_i)
\right]
+
 \sum_{i=1}^m\int_{a_i}^{a_{i+1}}
\left(\phi(x)- \frac{\phi_i}{a_{i+1}-a_{i}}   \right)^2 dx
\]
since
\[
\mathbf{E}\!\left[ (\hat \phi_i - \phi_i)^2 \right]=(\mathbf{E}\hat \phi_i - \phi_i)^2
+
\operatorname{Var}(\hat \phi_i)= \big( \mathrm{Bias}\,\big[\hat{\phi_i} \big] \big)^2+ \operatorname{Var}(\hat \phi_i).
\]

Here, only the first term is affected by regularization. The second term defines the bias associated with binning and represents the minimal bias inherent in the step-function representation of the unfolded distribution.

\subsection{Variance of ISE}
The variability of the  estimation of ISE can be defined as
\[
\mathrm{VISE}=\mathrm{Var}(\mathrm{ISE}) =
\mathbf{E}\!\left[ \mathrm{ISE}^2 \right]
- \Big( \mathbf{E}[\mathrm{ISE}] \Big)^2
\]
An algorithm with the lowest value of VISE  is preferable because it provides  a more stable solution  to the unfolding problem.
\subsection{Minimal Condition number}
Another measure of the quality of the unfolding process is the condition number of the correlation matrix for $\hat{\phi_i}/( a_{i+1}-a_{i}) $.  
The estimators of the probabilities $\hat{\phi_i}$  satisfy the  equation 
\begin{equation}
 \sum_{i=1}^m \hat{\phi_i}=1 
\end{equation}
and, as a result, the  correlation matrix is often nearly  singular. 
To address this, it is advisable to determine the minimum condition number (MCN) of the correlation matrix when excluding one bin from consideration.
\begin{equation}
\mathrm{MCN} =  \underset{k}{\text{argmin}}[ \mathrm{COND} (C_{-k})] ,
\end{equation}
where $C_{-k}$ is full correlation matrix with bin $k$ removed.
The condition number of correlation matrix measures the numerical stability of unfolding procedure and its sensitivity to small perturbations,
In particular, an algorithm with the lowest MCN value is preferable.
\subsection{Mean Square Error}
A popular measure of the accuracy of the unfolded distribution   \cite{zhig}  is the Mean Squared Error  ($\mathrm{MSE}$):
\begin{equation}
 \mathrm{MSE} =  \frac{1}{m} \sum_{i=1}^m   \mathbf{E}  \left ( \hat{\phi_i}- \phi_i\right )^2=  \frac{1}{m}  \sum_{i=1}^m
\left[ \big( \mathrm{Bias}\,\big[\hat{\phi_i} \big] \big)^2+ \operatorname{Var}(\hat \phi_i)
\right]
\end{equation}

The main disadvantage of MSE is that it cannot be used to compare the accuracy of two unfolded distributions obtained with different binning schemes. 
However, such a comparison is important when selecting an optimal binning scheme for an unfolding procedure.
\subsection{Coverage probability}
The coverage probability  \cite{kuus} is defined according as
\begin{equation}
 P_{\mathrm{cov}} =  \frac{1}{m} \sum_{i=1}^{m}
{P}\!\left(
\hat{\phi}_i - \hat{\sigma}_{\hat{\phi}_i }
< \phi_i <
\hat{\phi}_i + \hat{\sigma}_{\hat{\phi}_i }
\right),
\end{equation}
where
 $\hat{\sigma}_{{\hat{\phi}_i}}$ is  estimated uncertainty (e.g., standard deviation) of $\hat{\phi}_i$.
The main disadvantage of  $P_{\mathrm{cov}}$ is that it also cannot be used to compare the accuracy of two unfolded distributions obtained with different binning schemes. Such a comparison is important when selecting a better binning scheme for an unfolding procedure.
\subsection{Post-resolution}
Post-resolution, if can be estimated, provides information about the improvement of the effective resolution function compared with the intrinsic resolution of the experimental setup. An example of estimating the post-resolution function can be found in  \cite{zhig, Schmelling_2023}.
\vspace{0.5cm}
 
\section{Factors and parameters influencing  internal and external quality criteria. }

Different factors and parameters of the unfolding algorithm influence the quality of the unfolded distribution, as defined by the quality assessment methods presented in the previous section.These factors and parameters are listed below: 
\begin{enumerate}
\item Whether the measurement system is linear or nonlinear.
\item The generated  probability  distribution function $\phi^s(x)$ used to  calculate the  response matrix $R$,  and whether it is close to the true distribution  $\phi (x)$
\item The method used to calculate the response matrix $R$ (system identification), It can be:
\begin{itemize}
 \item The traditional method  \cite{review}, which may  lead to bias in unfolding result,  if the generated distribution used in the simulation, $\phi^s(x)$,  does not coincide with the true distribution  $\phi(x)$.
\item Methods used splitting bins on several sub bins and find  matrix element as average value of matrix elements of sub bins
\begin{itemize}
\item Approach used mean square method  presented in  \cite{ng}.
\item Inverse variance weighting method \cite{cochran} is considering  in next section
\end {itemize}
\end {itemize}
\item The number of simulated events  $k$ used to create the  mathematical model of measurement system.
\item The number of experimental events $n$. 
\item The number of bins $l$ in the experimental (measured) distribution.
\item The number of bins  $m$ in the unfolded (true) distribution
\item The type of binning:
\begin{itemize}
\item Equidistant binning. 
\item Non-equidistant binning, which can be defined using the k-means clustering method \cite{cluster}  and, in the multidimensional case, by the Voronoi method \cite{voronoi} additionally.
\end {itemize}
\item The regularization parameter of unfolding algorithm.  In Richardson-Lucy method \cite{Richardson,Lucy}  is the number of iterations. 
\item The initial guess for the unfolded distribution. This  is particularly important for Richardson-Lucy method in the case of small experimental statistics.
\end {enumerate}

The MISE, $\mathrm{Var}(\mathrm{ISE})$ and MCN  can be used as internal  quality criteria of  to compare different unfolding algorithm under variation of factors and parameters on which they depend. This approach was first time used in \cite{ng, Gagunashvili_2026}. 

 The use  MSN and  $P_{\mathrm{cov}}$   as internal  quality criteria  is rather restrictive. As shown  in \cite{review} it is possible to compare different algorithms when the same equidistant binning is used.  However, it is not possible to compare two different algorithm that present unfolded distributions with  two different binning schemes,  as was done in \cite {Gagunashvili_2026}. 

\section{Inverse variance weighting method for estimation the response matrix $R$}
 
Let the distribution of the generated data be represented by a histogram with $n_1, n_2, \ldots, n_m$ events in $m$ bins, and let the reconstructed data distribution be represented by a histogram with $n^r_1, n^r_2, \ldots, n^r_l$ events in $l$ bins. Note that the last bin represents the number of events lost due to the inefficiency of event registration.

The events in each bin $j$ of the generated histogram can be represented by a histogram of their reconstructed parameter, with $n^r_{1j}, n^r_{2j}, \ldots, n^r_{lj}$ events in the corresponding reconstructed bins. The elements of the response matrix $R$ are traditionally estimated as
\begin{equation}
r_{ij} = \frac{n^r_{ij}}{n_j}.
\end{equation}

The method assumes that the simulated PDF $\phi^s(x)$ is close to the unknown true PDF $\phi(x)$. This dependence on the simulated PDF $\phi^s(x)$ is particularly pronounced in the case of wide bins, as highlighted by \cite{kuus2}. Although this issue is less severe for narrow bins, the singularity problem becomes more pronounced, making the unfolding process more dependent on the strength of the regularization.

To mitigate this problem, let us introduce a subbinning scheme for each bin $j$ of the true distribution. The corresponding subhistogram consists of $s$ subbins, containing $n_{j1}, n_{j2}, \ldots, n_{js}$ events, respectively. The events in each subbin $jk$ are represented by a reconstructed histogram with $n^r_{1jk}, n^r_{2jk}, \ldots, n^r_{ljk}$ events in the corresponding reconstructed bins. We define

\begin{equation}
r_{ijk} = \frac{n^r_{ijk}}{n_{jk}}.
\end{equation}

The estimate of the response matrix element $r_{ij}$ is then obtained as the inverse-variance-weighted average

\begin{equation}
r_{ij} =
\frac{\displaystyle\sum_{k=1}^{s} w_{ijk} r_{ijk}}
{\displaystyle\sum_{k=1}^{s} w_{ijk}},
\end{equation}
where the weights are defined as

\begin{equation}
w_{ijk} =
\frac{1}{\mathrm{Var}(r_{ijk})},
\end{equation}

and the variance is calculated assuming a binomial distribution:

\begin{equation}
\mathrm{Var}(r_{ijk}) =
\frac{r_{ijk}(1-r_{ijk})}{n_{jk}}.
\end{equation}

This procedure minimizes the variance of each estimated response-matrix element, which is particularly important when the number of simulated events is relatively small.
\section{Methods used for the comparative study of the influence of binning choices on unfolding results}
Two multidimensional unfolding methods based on the optimization of a likelihood function are compared:

\begin{itemize}

\item{The Richardson--Lucy deconvolution method (RL) is an iterative method \cite{Richardson,Lucy}, in which the number of iterations is used as a regularization parameter.}

\item{The new data unfolding method with Mean Integrated Squared Error (MISE)  optimization (NG), proposed by the author of this  paper \cite{ng}, is a method based on entropy regularization.}

\end{itemize}

For both investigated unfolding methods, two binning schemes are considered:

\begin{itemize}

\item{K-means clustering is used for both the true and reconstructed histograms, with different numbers of bins.}

\item{Equidistant binning is used for the true distribution, while a clustering method is used to define the bins of the reconstructed histograms.}

\end{itemize}

The response matrix $R$ is calculated using simulated data generated according to a PDF $\phi^s(x)$ that differs from the unknown true PDF $\phi(x)$.

The inverse-variance weighting method described above is used to estimate the elements of the response matrix.

The dependence of the minimum MISE value on the regularization parameter is investigated for different numbers of bins in the unfolded distribution, $m$, and the measured distribution, $l$. The VISE and  MSN corresponding to the optimal MISE are investigated over the same range of binning configurations.

For each case with a close-to-optimal number of bins and regularization parameter, average unfolded distribution and  correlation matrix are calculated.

\section{Numeric example}
Two type of data was  simulated it is data for application unfolding algorithm and data for identification system or calculation $R$ matrix.
\subsection{ Data for unfolding}
\begin{figure}[H]
\vspace*{-0.5cm}
	\centering
 	\begin{subfigure}{0.495\linewidth}
		\includegraphics[width=\linewidth]{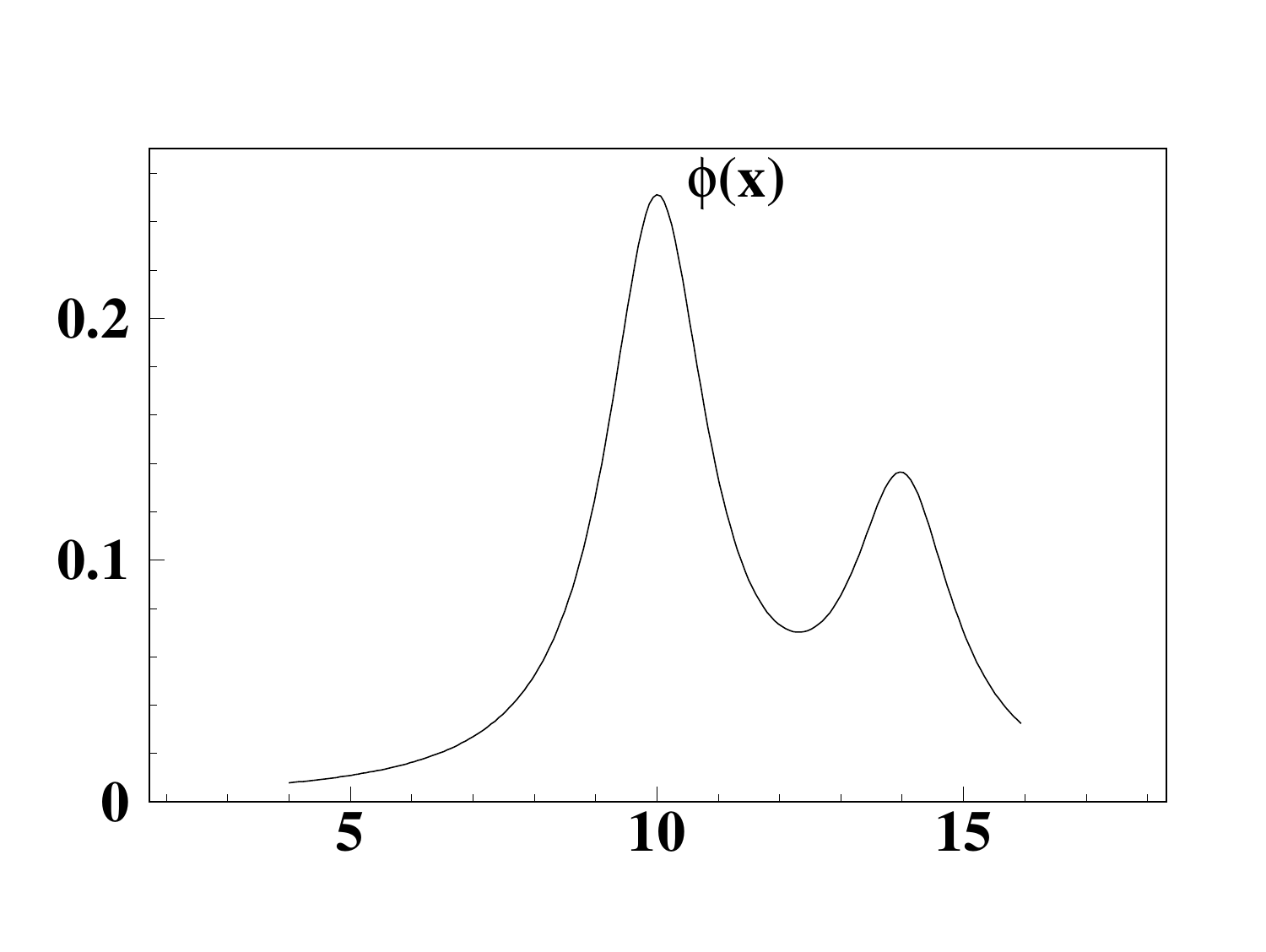}
	\end{subfigure}
   \hfill 
	\begin{subfigure}{0.495\linewidth}
		\includegraphics[width=\linewidth]{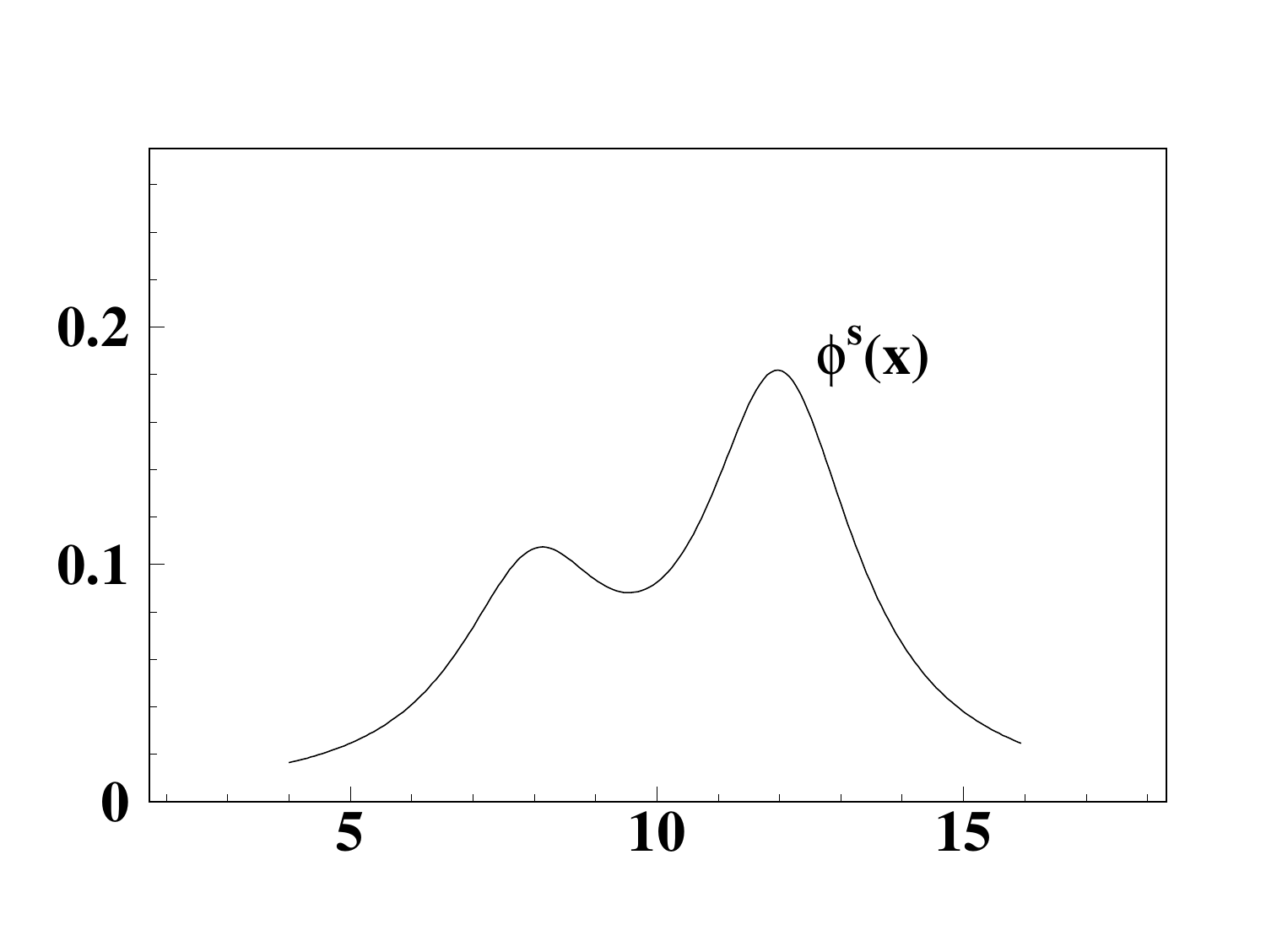}
	\end{subfigure}   
\vspace  *{-0.5cm}   
	\caption{True  PDF $\phi (x)$ (left), PDF used for simulation   $\phi^s(x)$  (right).}
	\label{fig:trueonl}
\end{figure}

\begin{figure}[H]
	\centering
	\begin{subfigure}{0.495\linewidth}
		\includegraphics[width=\linewidth]{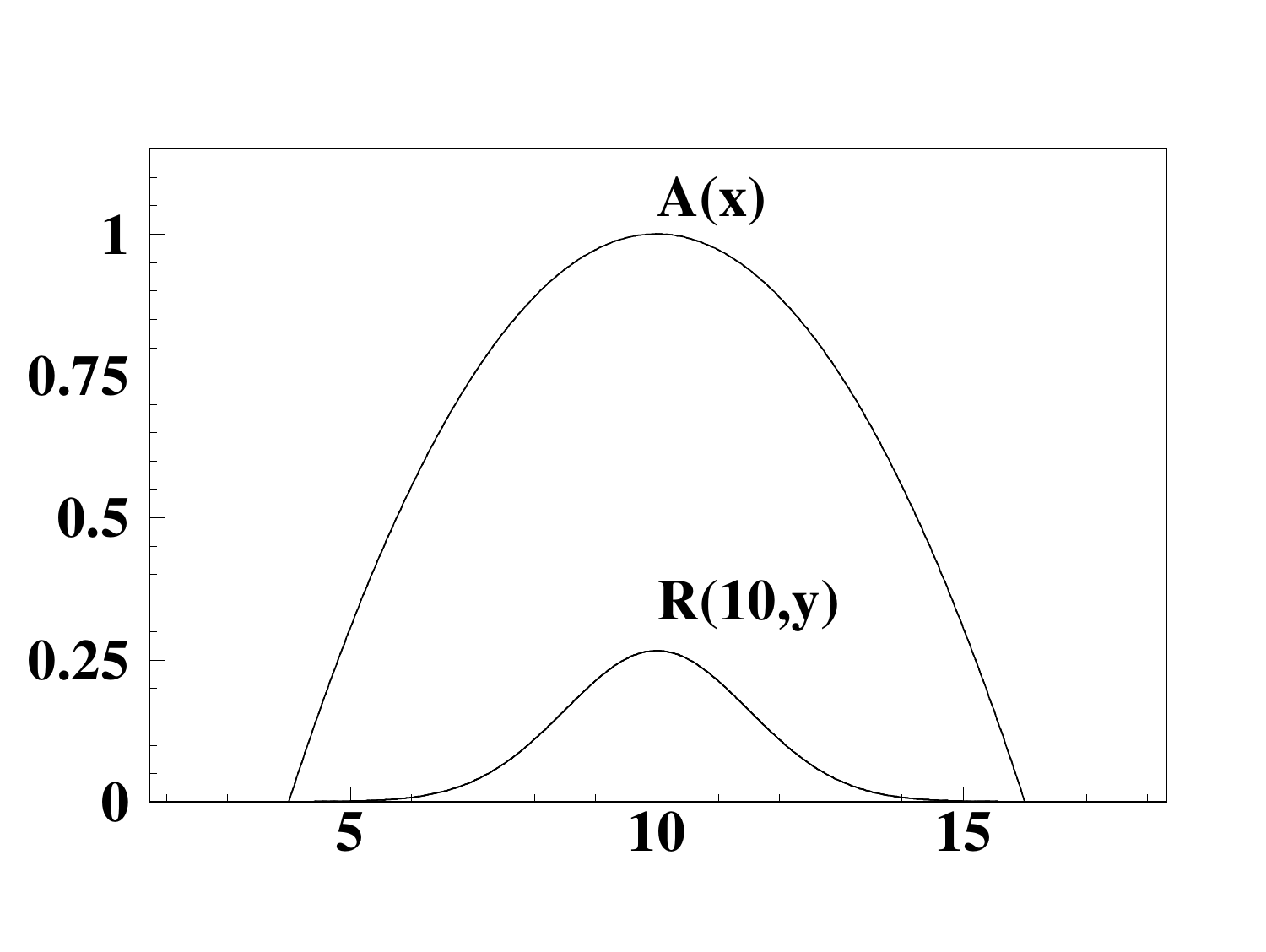}
	\end{subfigure}
\vspace  *{-0.5cm} 
	\caption{Acceptance $A(x)$\/ and resolution function $R(x,y)$,   for $x=10$ .} 
	\label{fig:accept}
\end{figure}

Following to \cite{zhig}  a true PDF $\phi(x)$  is described by a sum of two Breit--Wigner functions defined over interval $[4, 16]$,  as shown in Figure~\ref{fig:trueonl}(left):
\begin{equation}
 \phi(x) \propto 2\frac{1}{(x-10)^2+1} + \frac{1}{(x-14)^2+1},
\label{testform}
\end{equation}
from which the measured PDF $f(y)$  is obtained according to equation (\ref{fred})
with the acceptance function $A(x)$ (Figure~\ref{fig:accept}):
\begin{equation}
A(x)=1-\frac{(x-10)^2}{36}  , \label{acc}
\end{equation}
and the resolution function describing Gaussian smearing (Figure~\ref{fig:accept}):
\begin{equation}
R(x,y)=\frac{1}{\sqrt{2\pi}\sigma}\exp\left(-\frac{(y-x)^2}{2\sigma^2}\right), \, \sigma=1.5\; \label{res}  .
\end{equation}
Sample for analysis contain  10000  events simulated according formulas  (\ref{testform}) and \ref{fred}.
\subsection{Data for system identification}
To create model of  the measurement system, generated (PDF) presented  by a sum of two Breit-Wigner functions defined over interval $[4, 16]$,  as shown in Figure~\ref{fig:trueonl}(right):
\begin{equation}
 \phi^s (x) \propto \frac{1.5^2}{(x-8)^2+1.5^2} + 2\frac{1.5^2}{(x-12)^2+1.5^2}.  \label{rcs}  
\end{equation}
From this, the reconstructed  PDF $f^s(y)$  is obtained according to equation (\ref{fred}) with the same acceptance  $A(x)$  equation  (\ref{acc}) and resolution function  $R(x,y)$  (\ref{res}).

Sample for system identification  contain  1000000  events simulated according formulas  (\ref{rcs}) and \ref{fred}.
\subsection{Binning as a regularization factor in unfolding}
The dependence of the optimal MISE value and the corresponding VISE and MSN values on the number of bins was investigated. The measured distributions were presented using 20, 40, 60, 80, and 100 bins, while the unfolded distributions were presented using 15, 25, 35, and 40 bins.
Two types of binning were considered for the unfolded distributions: non-equidistant binning obtained using the k-means clustering method and equidistant binning. In both cases, the measured distributions were represented using non-equidistant binning.
The number of sub-bins within each bin of the unfolded distribution was fixed at 5.

A total of 500 pseudodata samples, each containing 10,000 events, were used to calculate the quality assessment metrics (MISE, VISE, and MSN).

\subsubsection{Unfolded PDF with non-equidistant binning}
The dependence of the unfolding quality metrics on the binning configuration was investigated for the case of non-equidistant binning scheme.   The results are presented graphically below. The marker \ding{115} denotes the case with 15 bins in the unfolded distribution, \ding{110} corresponds to 25 bins, \ding{108} corresponds to 35 bins, and \ding{72}  to 40 bins.
\begin{figure}[H]
	\centering
 	\begin{subfigure}{0.495\linewidth}
		\includegraphics[width=\linewidth]{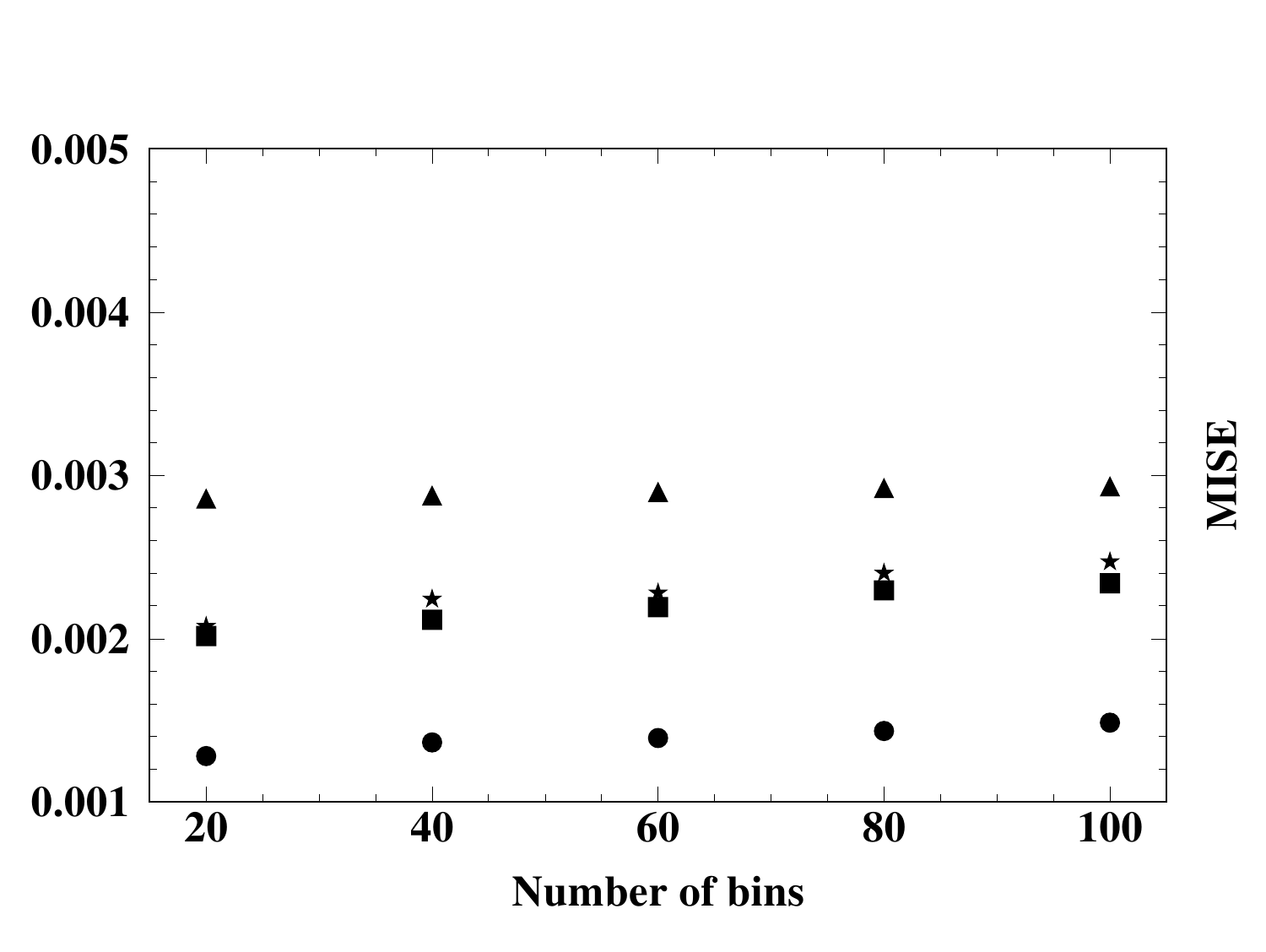}
	\end{subfigure}
   \hfill 
	\begin{subfigure}{0.495\linewidth}
		\includegraphics[width=\linewidth]{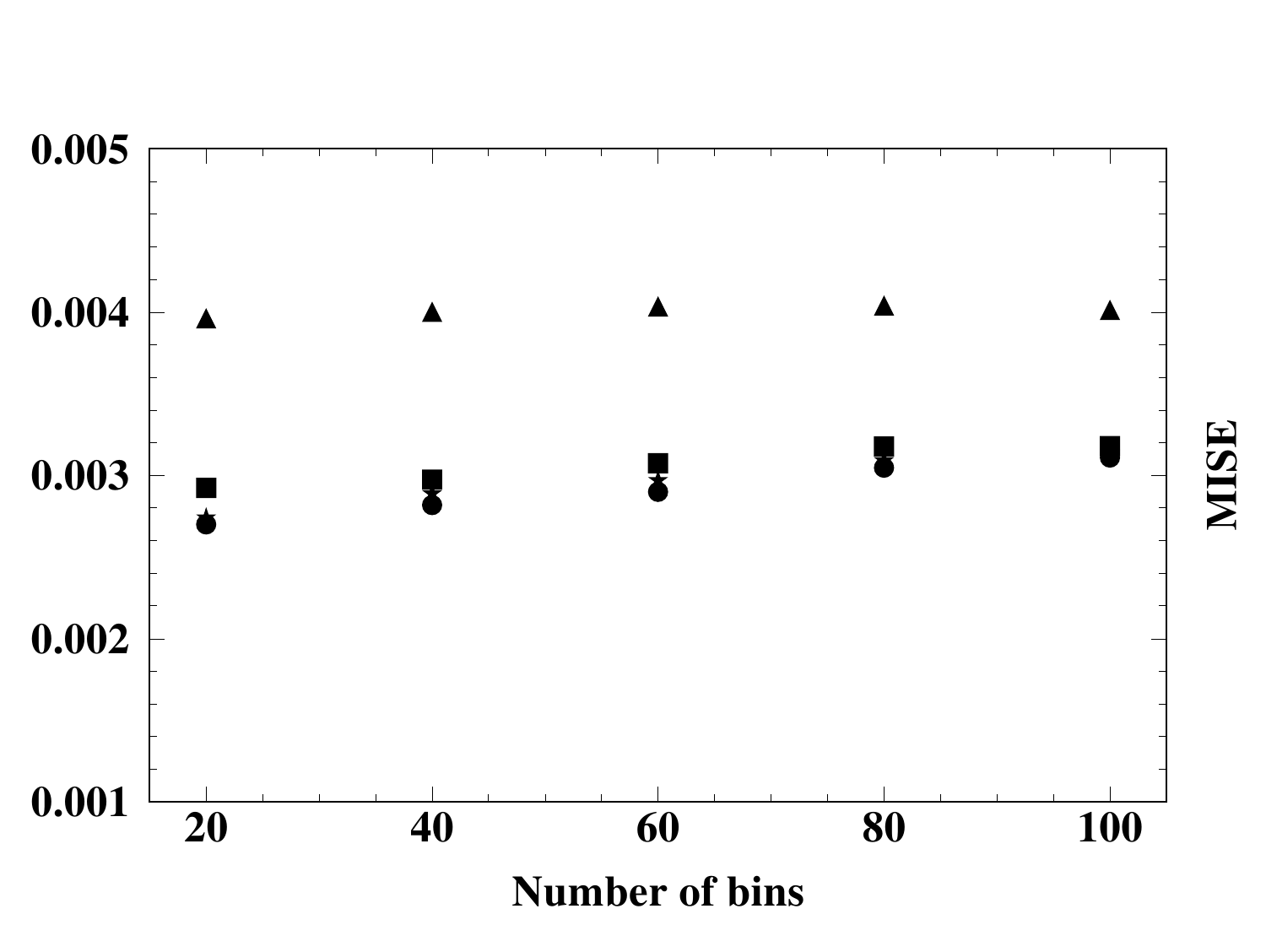}
	\end{subfigure}   
\vspace *{-0.3cm}    
	\caption{Minimal  value of MISE for NG (left)  and  for RL  method (right). }
	\label{fig:mise_cl}
\end{figure} 
Figure~\ref{fig:mise_cl} shows that the minimum MISE values are achieved with 35 bins in the unfolded distribution in both cases, with the NG method yielding a substantially lower MISE.

\begin{figure}[H]
	\centering
 	\begin{subfigure}{0.495\linewidth}
		\includegraphics[width=\linewidth]{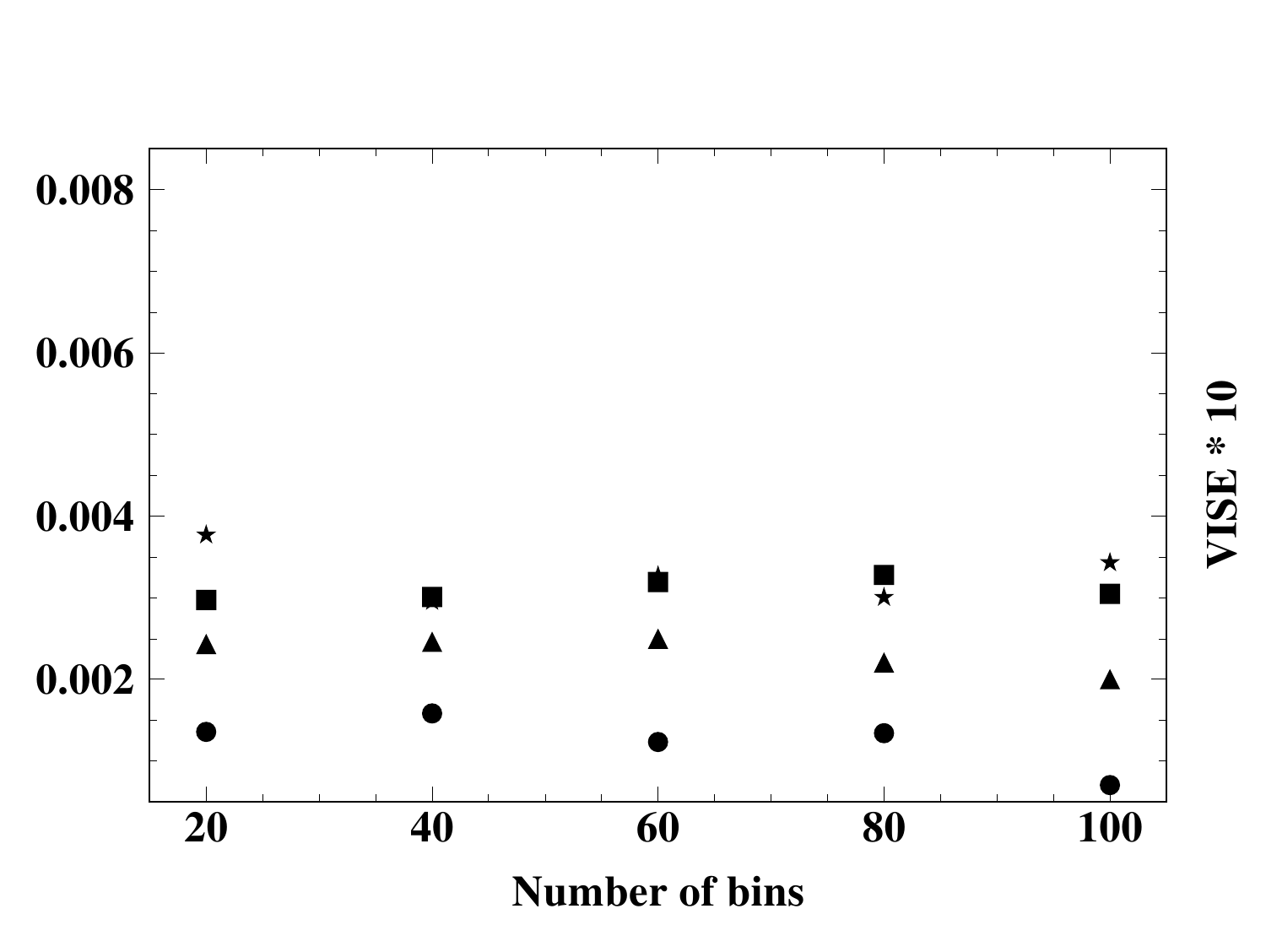}
	\end{subfigure}
   \hfill 
	\begin{subfigure}{0.495\linewidth}
		\includegraphics[width=\linewidth]{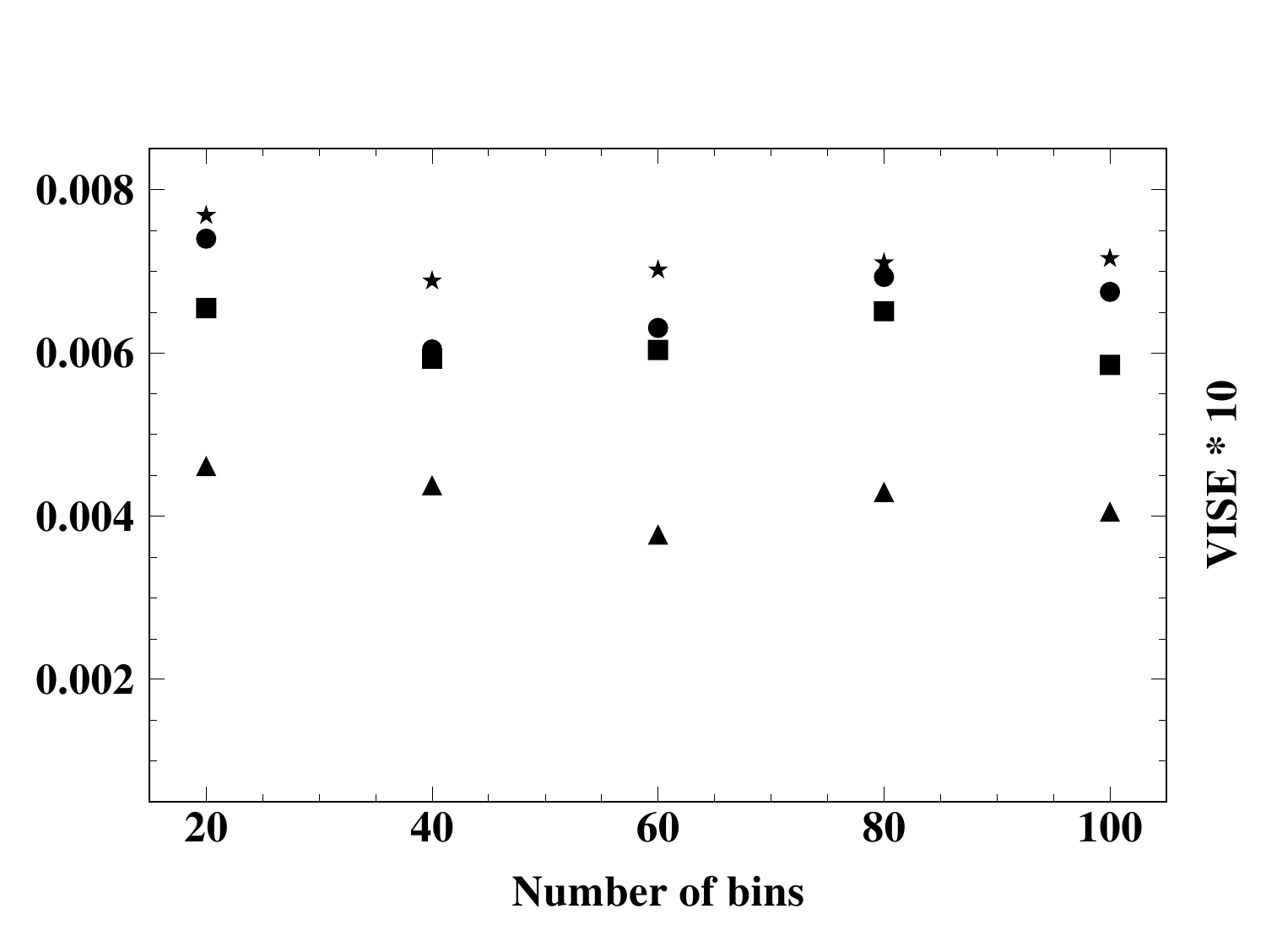}
	\end{subfigure}   
\vspace *{-0.3cm}    
	\caption{VISE for optimal value of MISE for case of NG method (left) and for RL method (right). }
	\label{fig:er_cl}
\end{figure} 
\vspace*{-0.25cm} 
Figure ~\ref{fig:er_cl} shows that the minimum VISE values at the optimal MISE are achieved with 35 bins in the unfolded distribution for the NG method and 15 bins for the RL method. The VISE value is substantially lower for the NG method.
\begin{figure}[H]
	\centering
 	\begin{subfigure}{0.495\linewidth}
		\includegraphics[width=\linewidth]{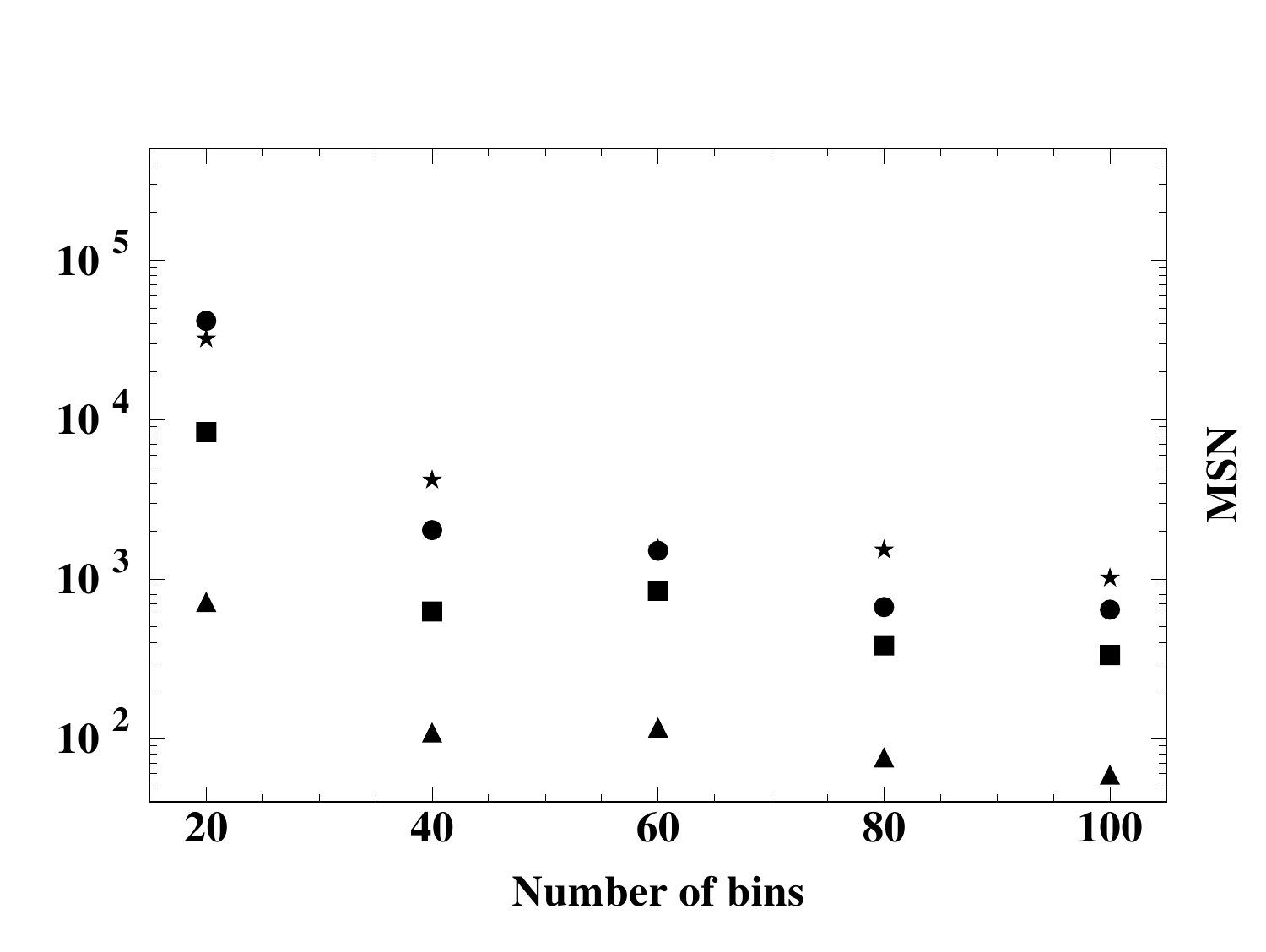}
	\end{subfigure}
   \hfill 
	\begin{subfigure}{0.495\linewidth}
		\includegraphics[width=\linewidth]{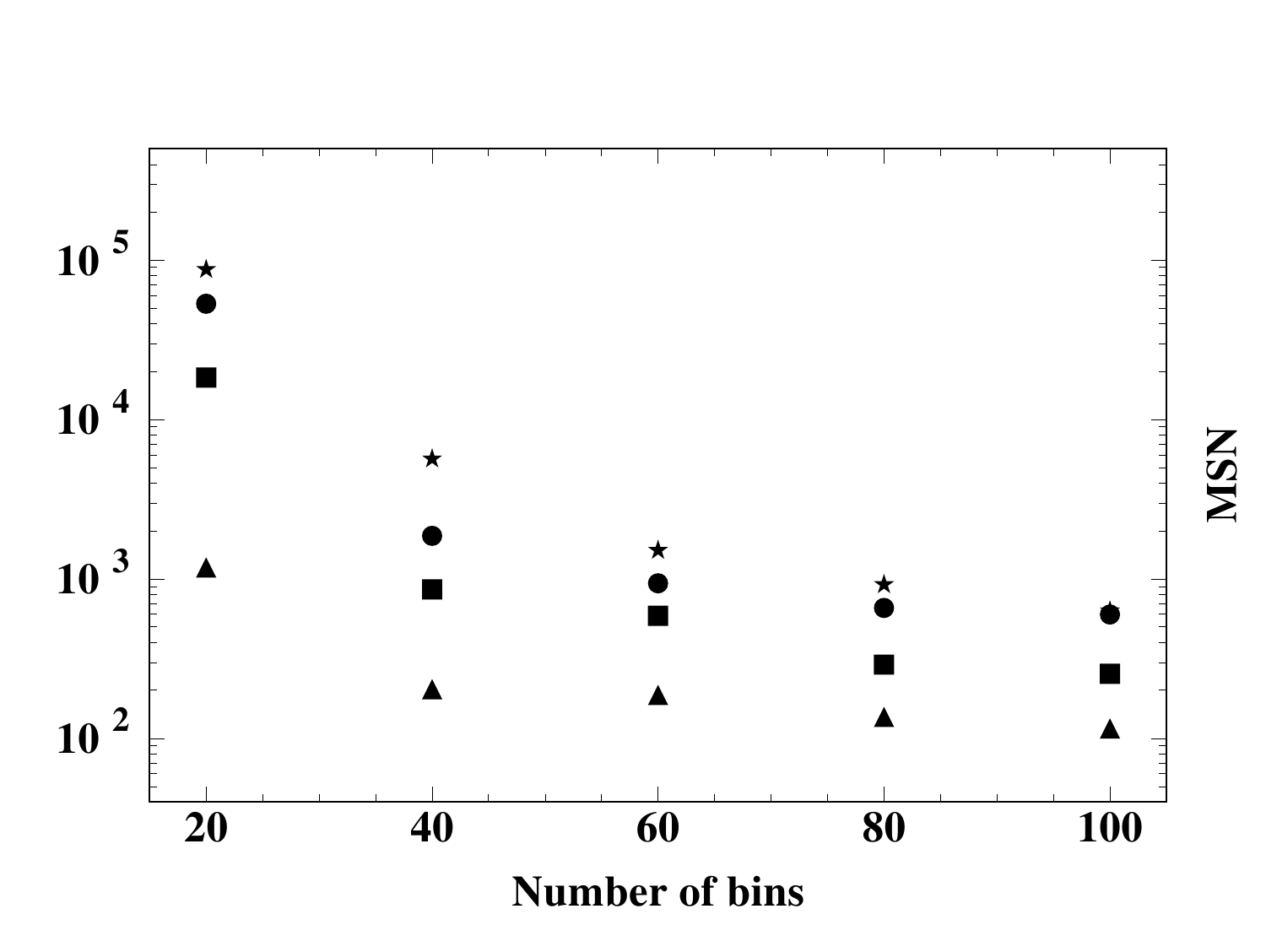}
	\end{subfigure}        
	\caption{MSN for optimal value of MISE for case of NG method (left) and for RL method (right). }
	\label{fig:msn_cl}
\end{figure} 
\vspace*{-0.25cm} 
Figure ~\ref{fig:msn_cl} shows that the MSN decreases as the number of bins in the measured distribution decreases to 100, at the optimal MISE, for both methods. The minimum MSN values are achieved with 15 bins in the unfolded distribution. In all cases, the MSN values obtained using the NG method are slightly lower than those obtained using the RL method.
\begin{figure}[H]
	\centering
 	\begin{subfigure}{0.495\linewidth}
		\includegraphics[width=\linewidth]{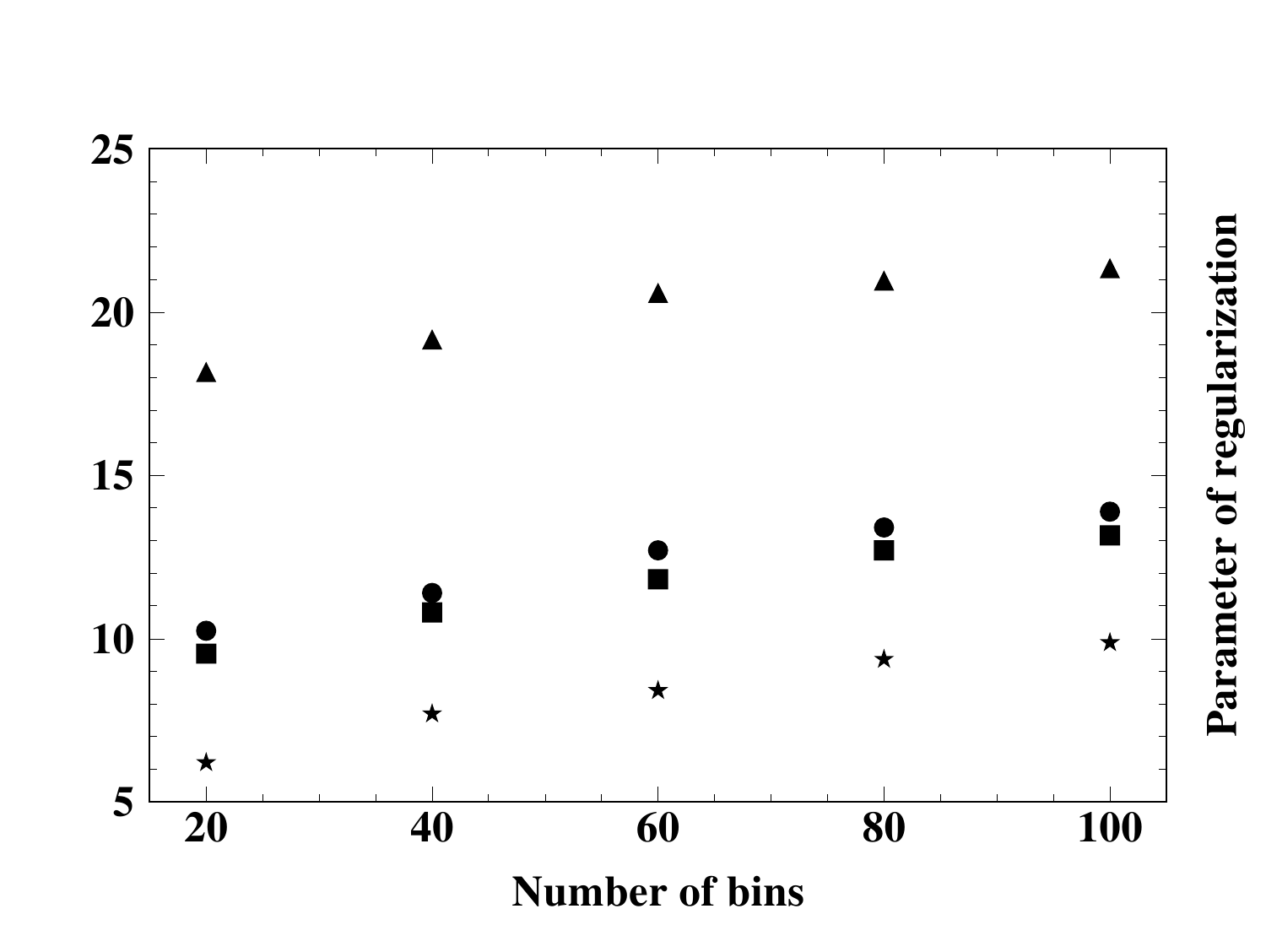}
	\end{subfigure}
   \hfill 
	\begin{subfigure}{0.495\linewidth}
		\includegraphics[width=\linewidth]{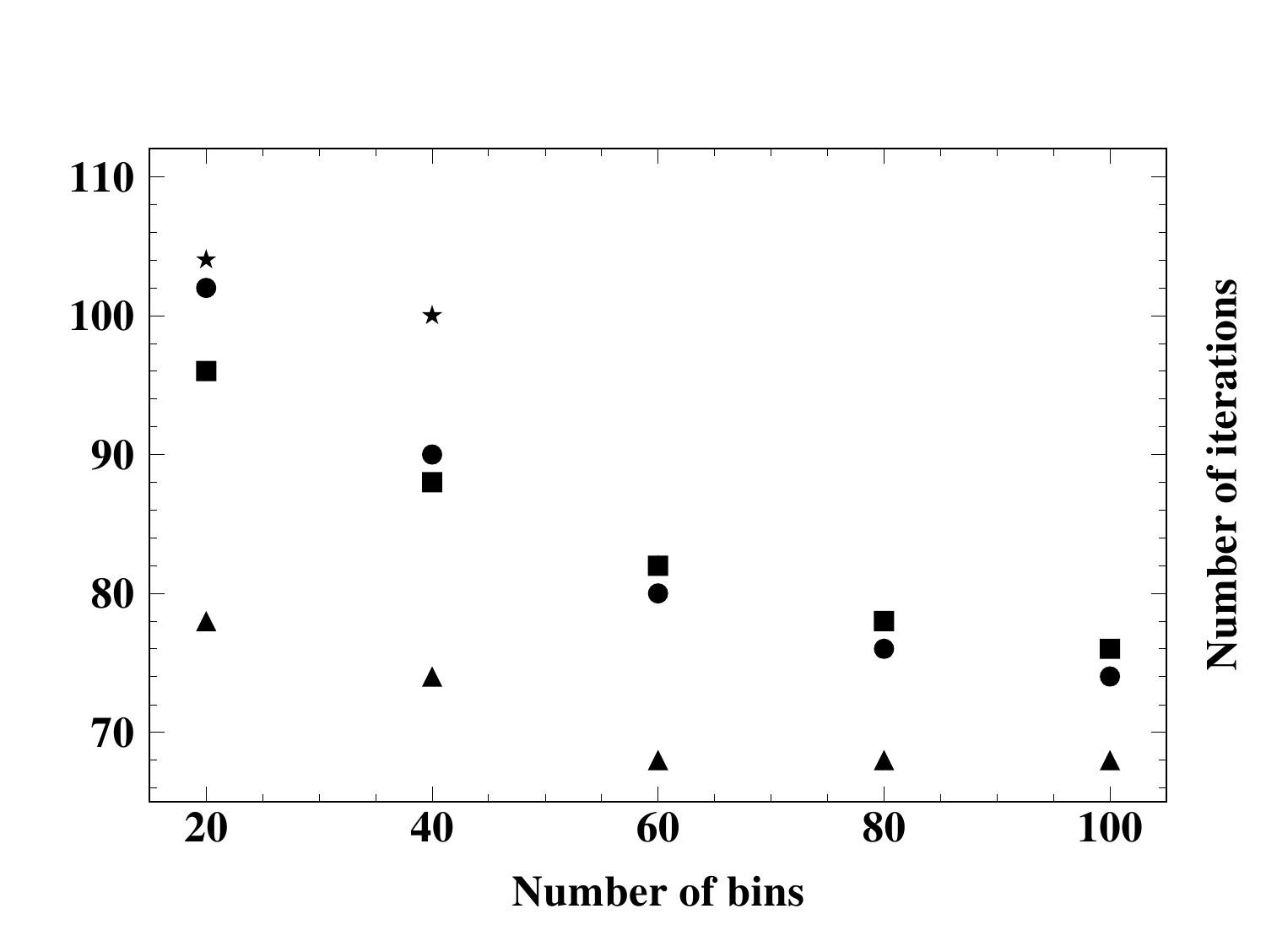}
	\end{subfigure}        
	\caption{Regularization parameters for optimal value of MISE for case of NG method (left) and for RL method (right). }
	\label{fig:param}
\end{figure} 
As shown in Figure~\ref{fig:param}, the regularization parameter for the NG method increases with the number of bins in the measured distribution, while the number of iterations for the RL method decreases.

Based on a combination of parameters used to assess the quality of the results, the NG method produces better results than the RL method.
\newpage
To continue analysis the average unfolded distribution and were calculated for the case  80 bins in the measured distribution and 35 bins for unfolded distribution.  

\begin{figure}[H]
	\centering
	\begin{subfigure}{0.495\linewidth}
		\includegraphics[width=\linewidth]{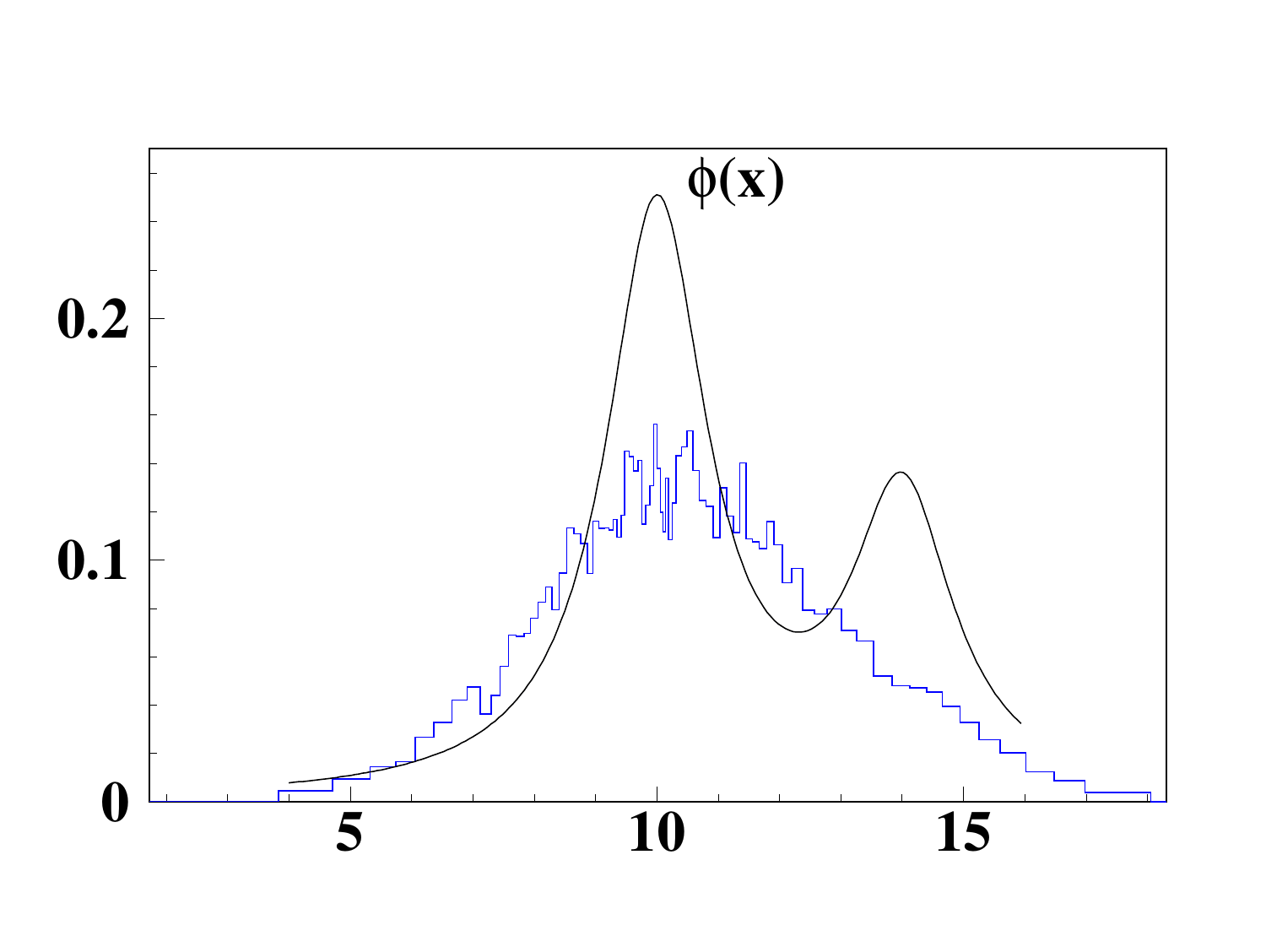}
	\end{subfigure}
\vspace  *{-0.5cm} 
	\caption{Measured distribution.} 
	\label{fig:true}
\end{figure}

 The measured distribution with 80 bins  is shown in Figure~\ref{fig:true} and is used for comparing both methods.

\begin{figure}[H]
	\centering
 	\begin{subfigure}{0.495\linewidth}
		\includegraphics[width=\linewidth]{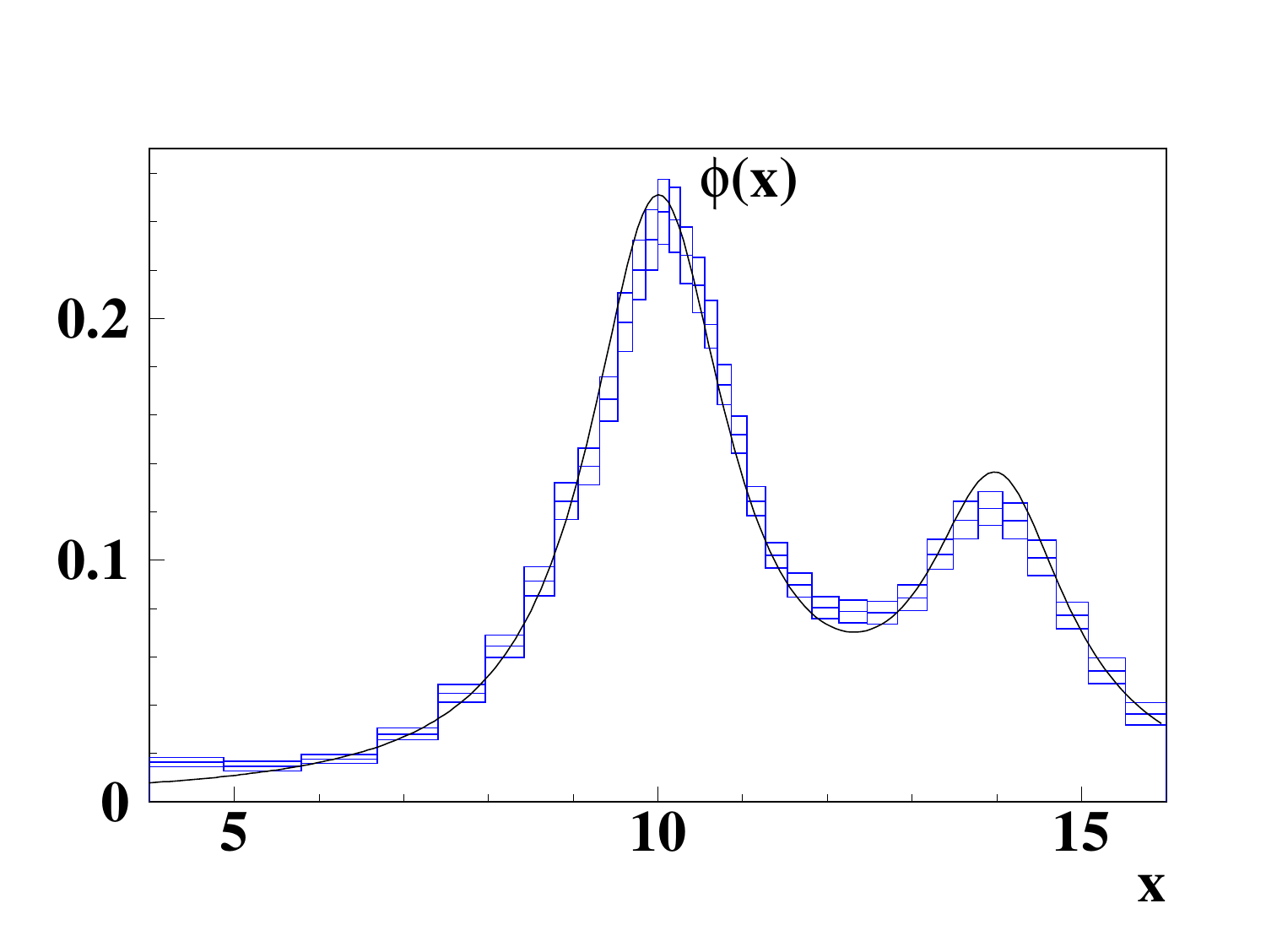}
	\end{subfigure}
   \hfill 
	\begin{subfigure}{0.495\linewidth}
		\includegraphics[width=\linewidth]{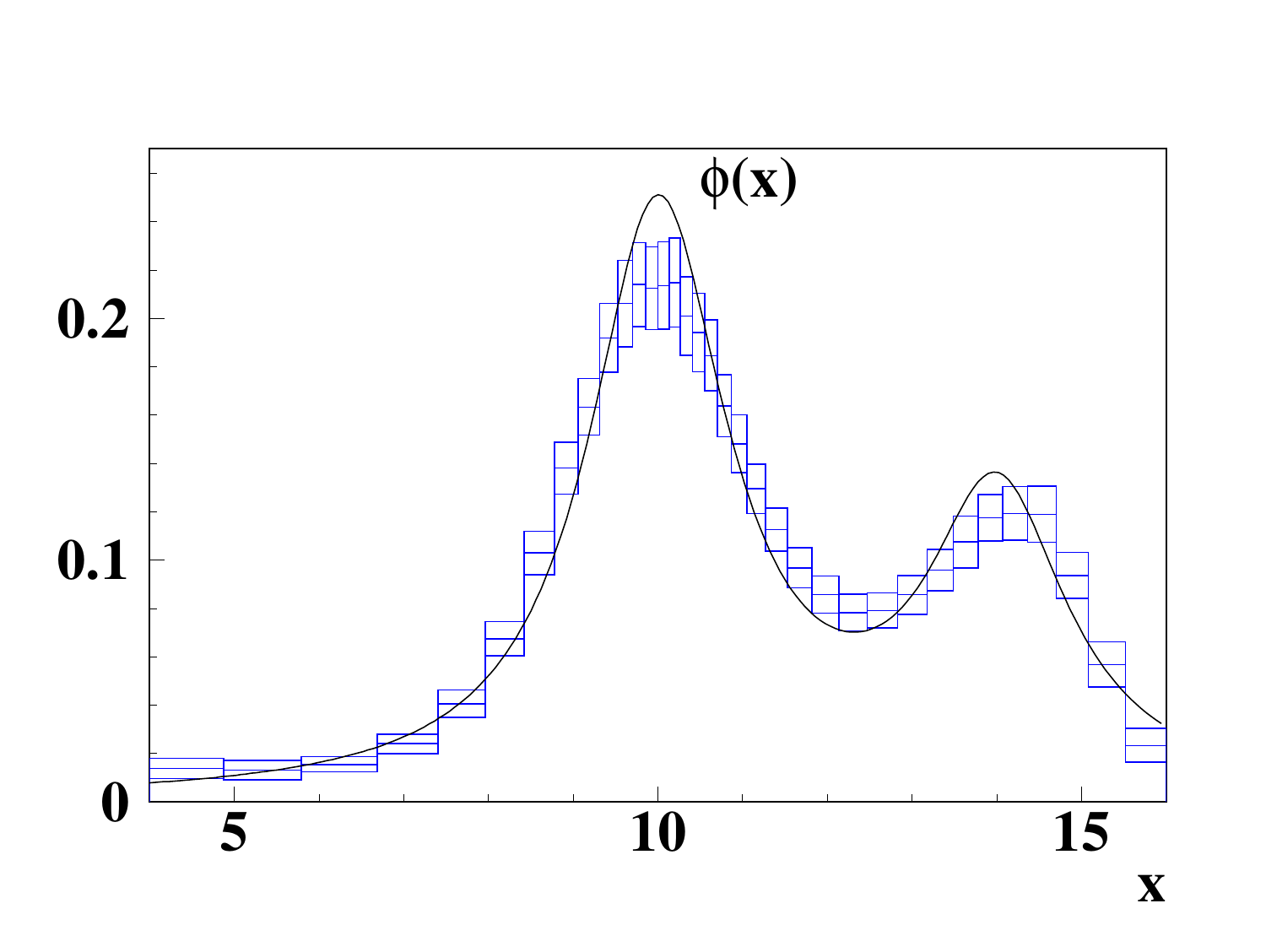}
	\end{subfigure}        
	\caption{Average unfolded PDF for NG method with the minimum MISE value (left) and  average unfolded PDF  for RL method with the minimum MISE value (right). }
	\label{fig:unfol}
\end{figure}

\begin{figure}[H]
	\centering
 	\begin{subfigure}{0.495\linewidth}
		\includegraphics[width=\linewidth]{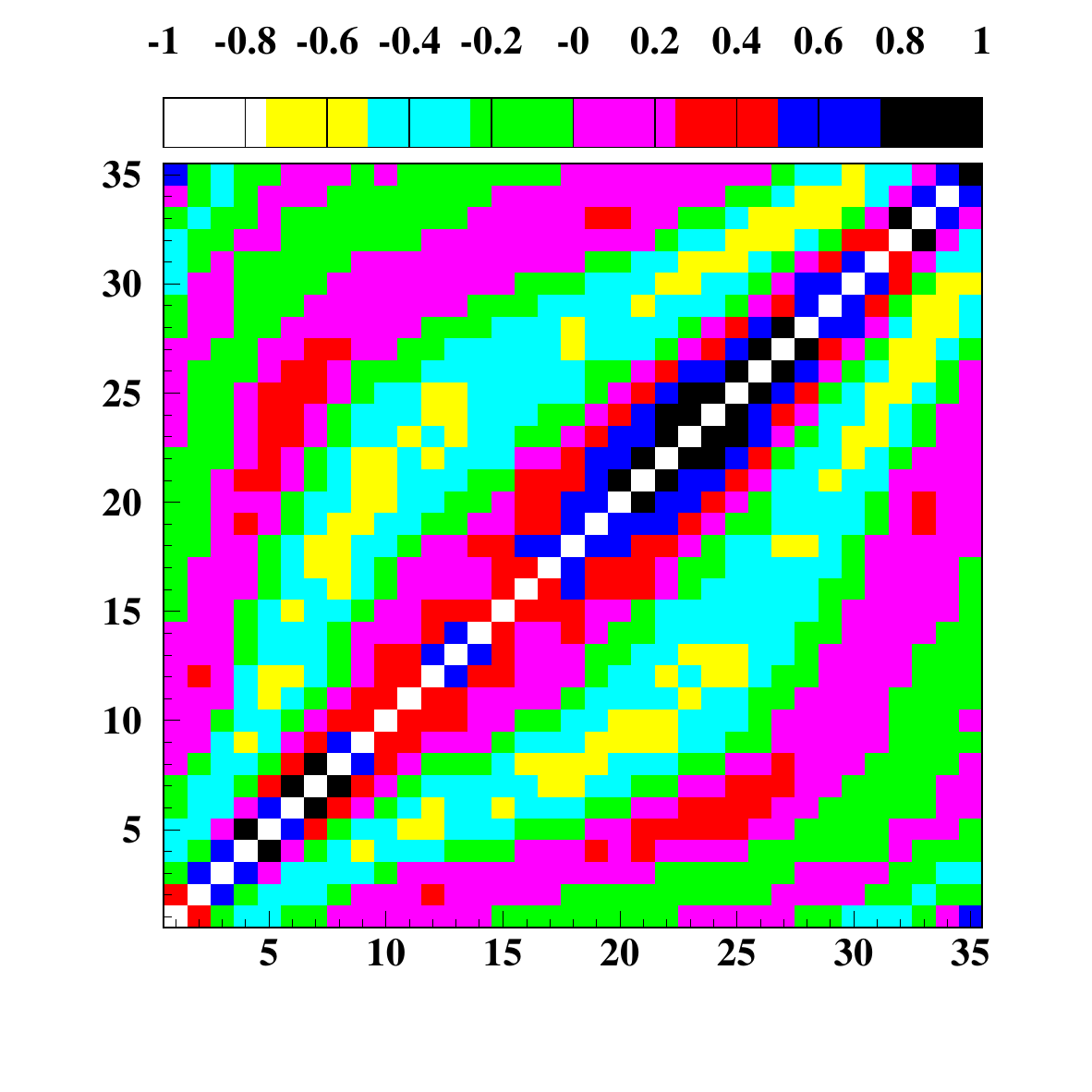}
	\end{subfigure}
   \hfill 
	\begin{subfigure}{0.495\linewidth}
		\includegraphics[width=\linewidth]{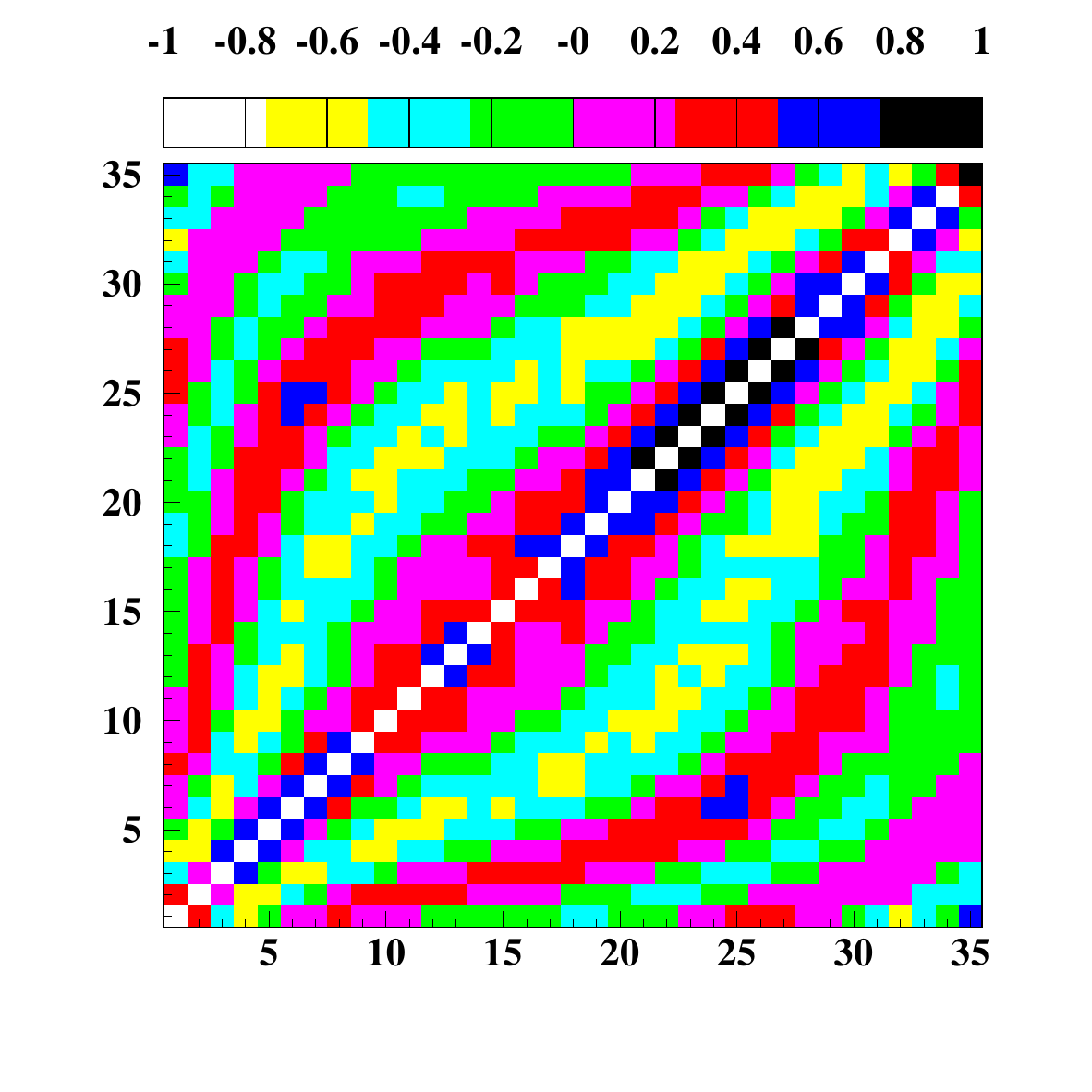}
	\end{subfigure}   
\vspace*{-0.5cm}   
	\caption{Correlation matrices for NG method  (left) and   for RL method (right)}
	\label{fig:corr}
\end{figure}

The average unfolded distribution with 35 bins  is shown in  Figure~\ref{fig:unfol}. 
The error bar represents the standard deviation of the estimate $\hat{\phi_i}/(a_{i+1}-a_{i})$, for bin $i$.
\vspace*{-0.8cm} 
 The corresponding correlation matrices for both methods are also presented in Figure~\ref{fig:corr}.
\newpage
\vspace*{-2.5cm}
In addition, a table summarizing the optimal MISE, VISE and MSN values, together with the corresponding regularization parameters, is provided that again demonstrate prevalence of new data unfolding with MISE optimization under Richardson-Lucy method.
\begin{table}[H]
\begin{center}
\captionof{table}{Characteristics of the optimal  MISE, VISE and  MCN for NG and RL algorithms  in case 35 non-equidistant bins for unfolded distribution and 80 bins for measured  one. }
\begin{tabular}{ | c|  c  c   c  c |   }
\hline
 Method &  MISE  & VISE &   MCN & Reg. par. \\
\hline 
  NG &  $1.432 \times 10^{-2}$  & $1.337 \times 10^{-4}$  & 665.0 &13.4 \\
  RL &  $3.047 \times 10^{-2}$ & $6.929 \times 10^{-4}$  & 655.6   &76\\
\hline
\end{tabular}
\end{center}
\end{table}
\subsubsection{ Unfolded PDF with equidistant binning}
The dependence of the unfolding quality metrics on the binning configuration was investigated for the case of equidistant binning scheme.   The results are presented graphically below. For all  cases measured distribution was presented in non-equidistant binning and notation of marker denoted different number of bins in unfolded distribution are the same as in previous subsection.  
\newpage
\vspace *{-1.5cm} 
\begin{figure}[H]
	\centering
 	\begin{subfigure}{0.495\linewidth}
		\includegraphics[width=\linewidth]{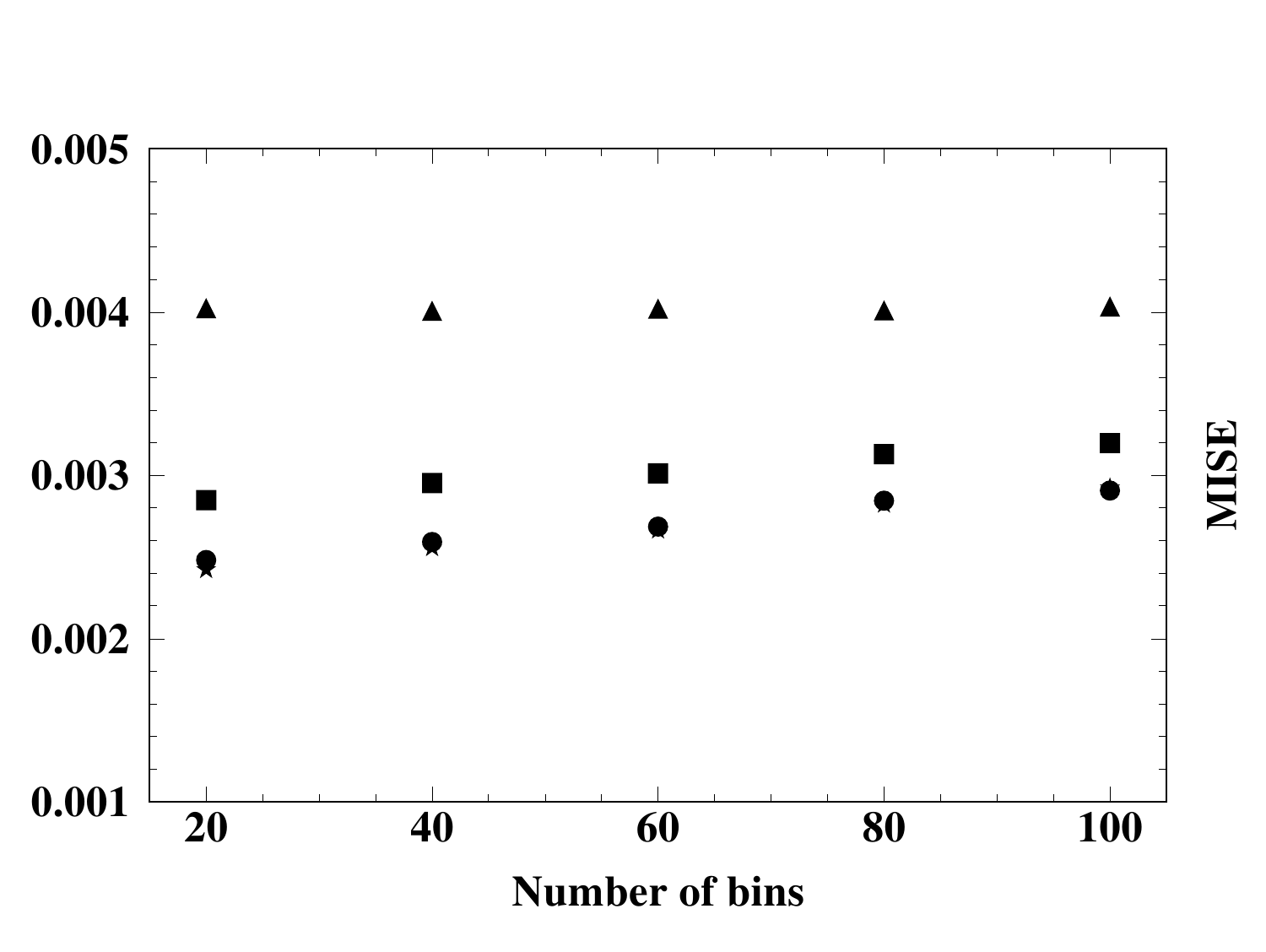}
	\end{subfigure}
   \hfill 
	\begin{subfigure}{0.495\linewidth}
		\includegraphics[width=\linewidth]{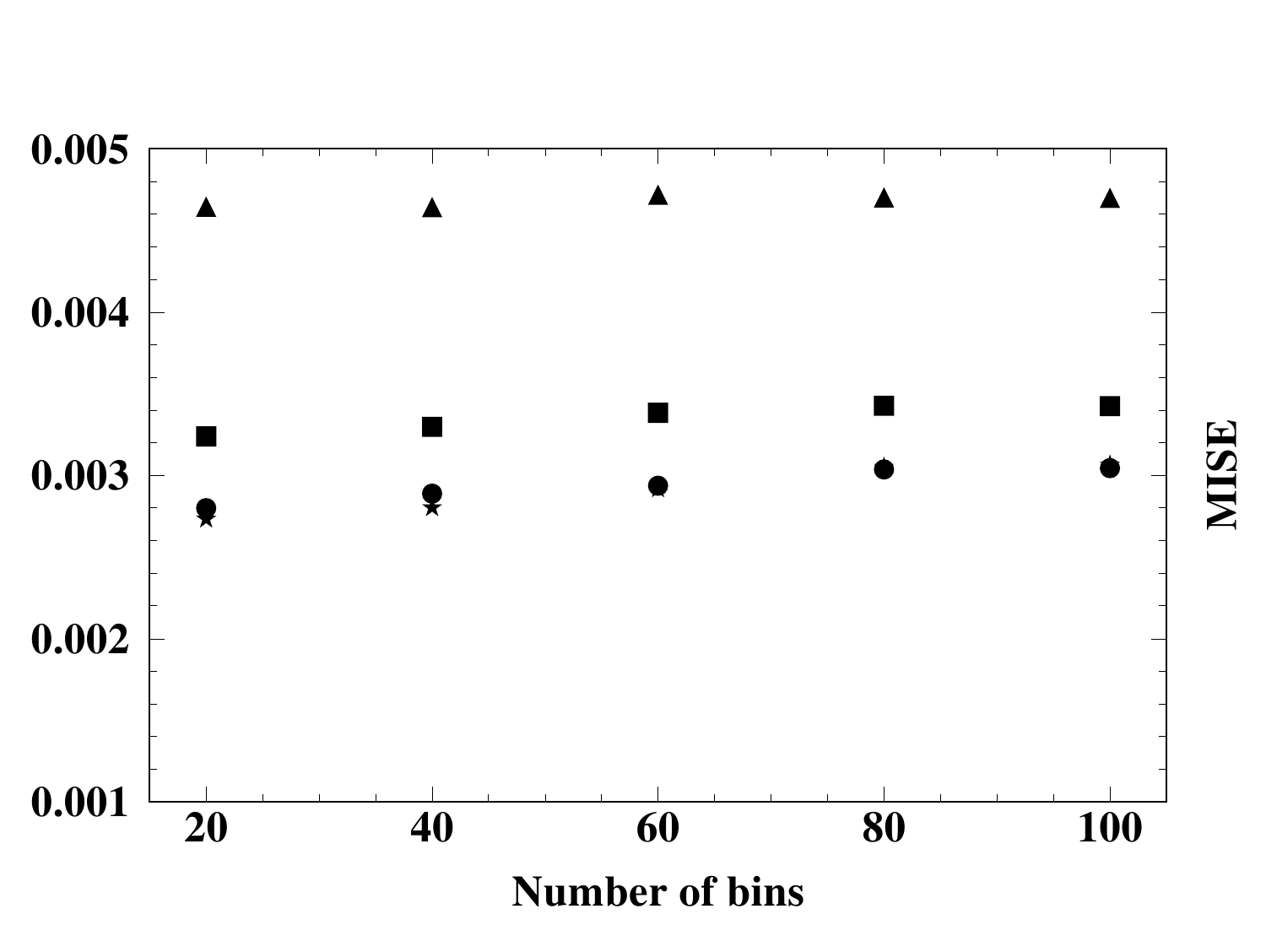}
	\end{subfigure}   
        \caption{Minimal  value of MISE for NG (left)  and  for RL  method (right). }
	\label{fig:mise_eq}
\end{figure} 
Figure~\ref{fig:mise_eq} shows that the minimum MISE values are achieved approximately  with 35 bins in the unfolded distribution in both cases, with the NG method yielding a  slightly lower MISE.
\begin{figure}[H]
	\centering
 	\begin{subfigure}{0.495\linewidth}
		\includegraphics[width=\linewidth]{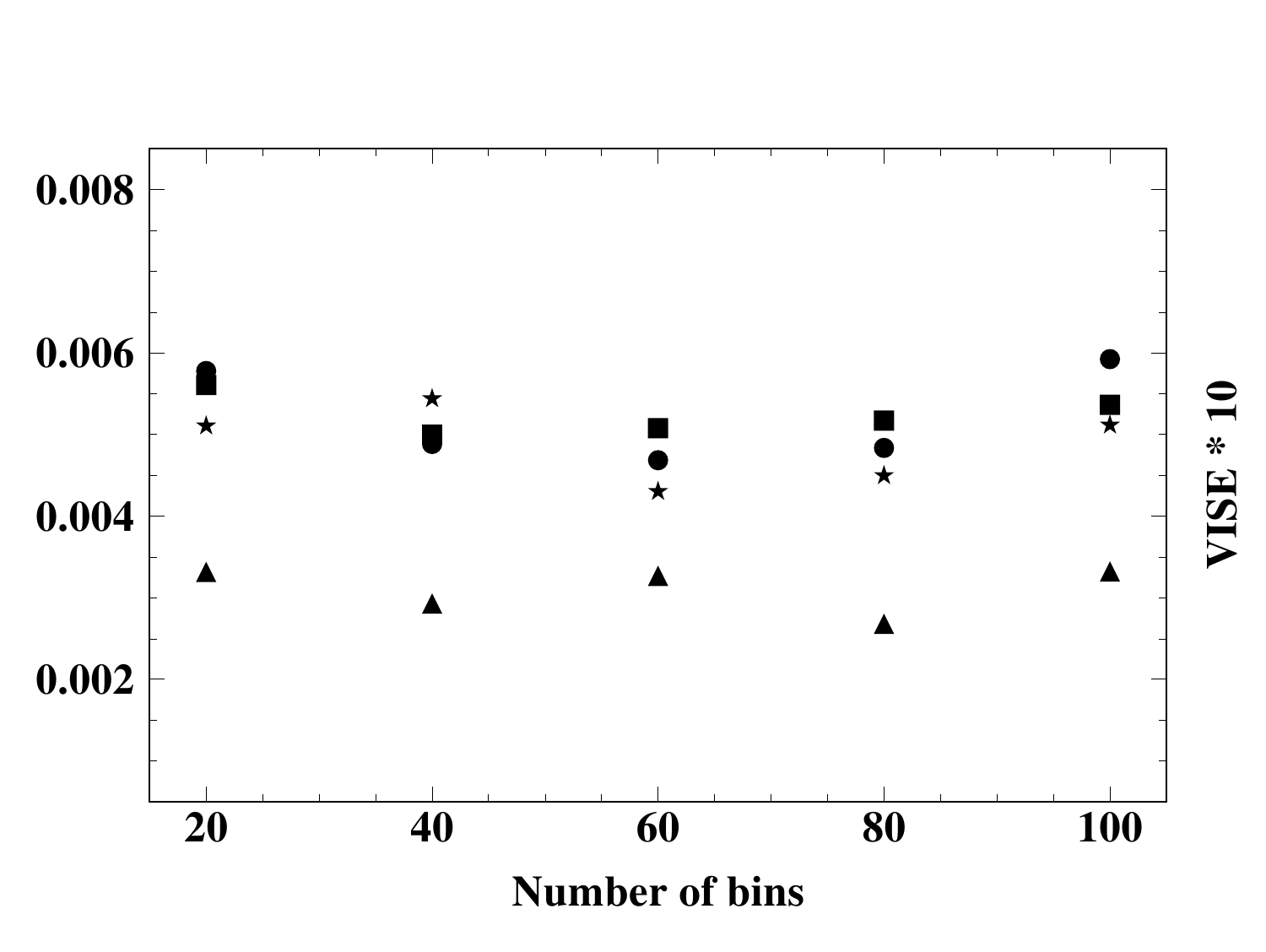}
	\end{subfigure}
   \hfill 
	\begin{subfigure}{0.495\linewidth}
		\includegraphics[width=\linewidth]{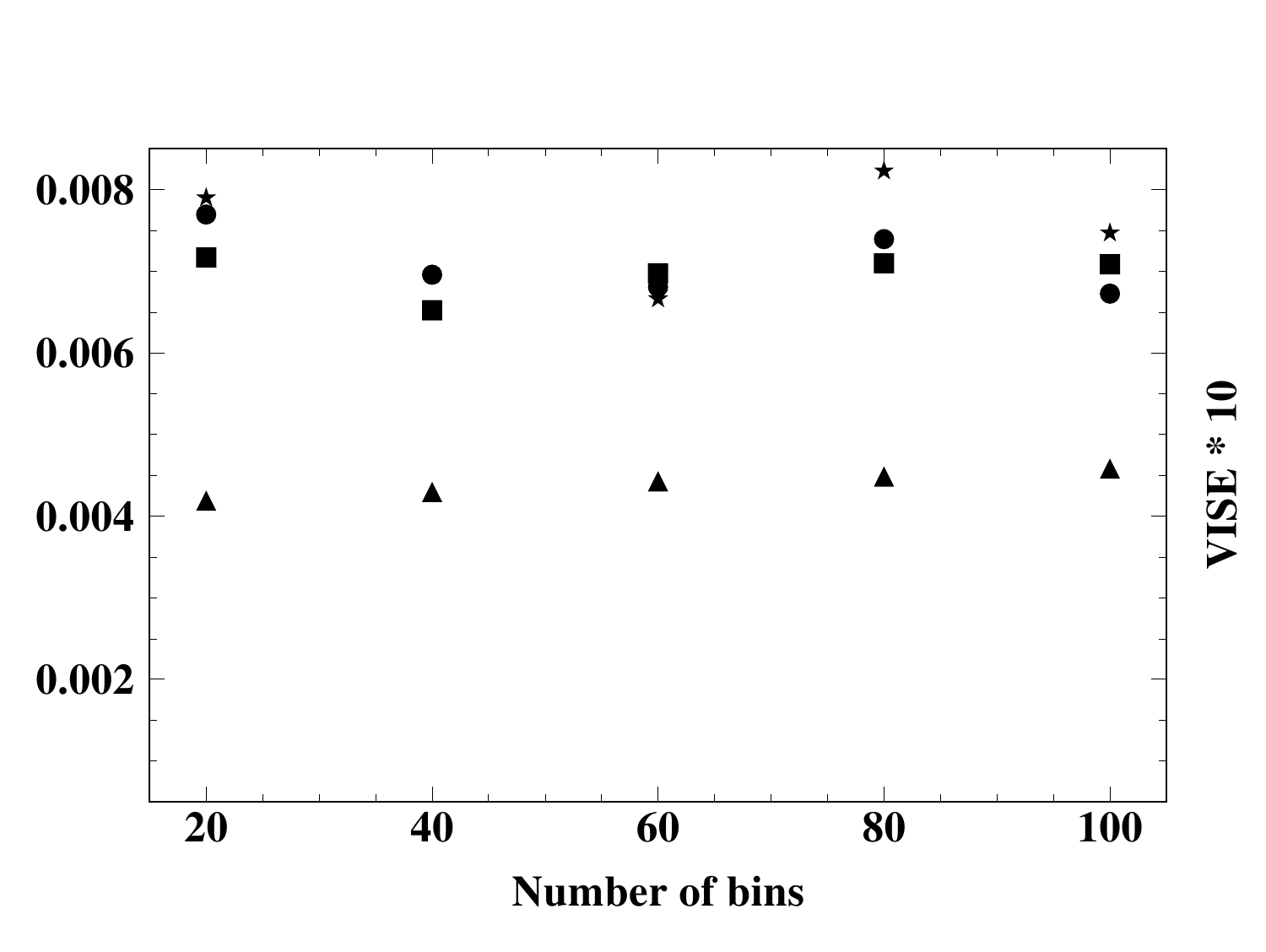}
	\end{subfigure}   
	\caption{VISE for optimal value of MISE for case of NG method (left) and for RL method (right).}
	\label{fig:er_eq}
\end{figure} 
Figure ~\ref{fig:er_eq} shows that the minimum VISE values at the optimal MISE are achieved with 15 bins in the unfolded distribution for the NG method and 15 bins for the RL method. The VISE value for 35 bins in the unfolded distribution lower for the NG method.
\newpage
\vspace *{-0.5cm} 
\begin{figure}[H]
	\centering
 	\begin{subfigure}{0.495\linewidth}
		\includegraphics[width=\linewidth]{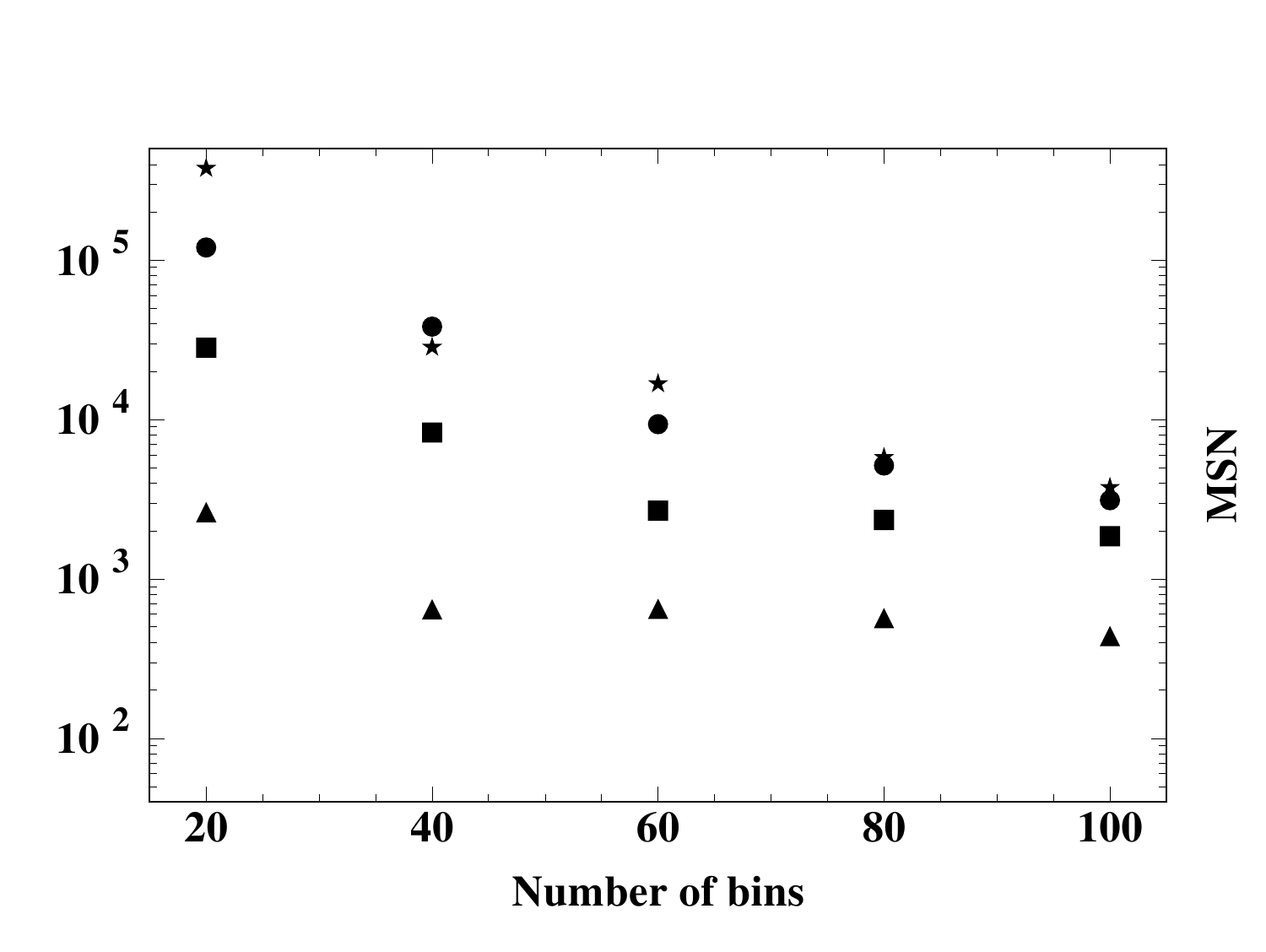}
	\end{subfigure}
   \hfill 
	\begin{subfigure}{0.495\linewidth}
		\includegraphics[width=\linewidth]{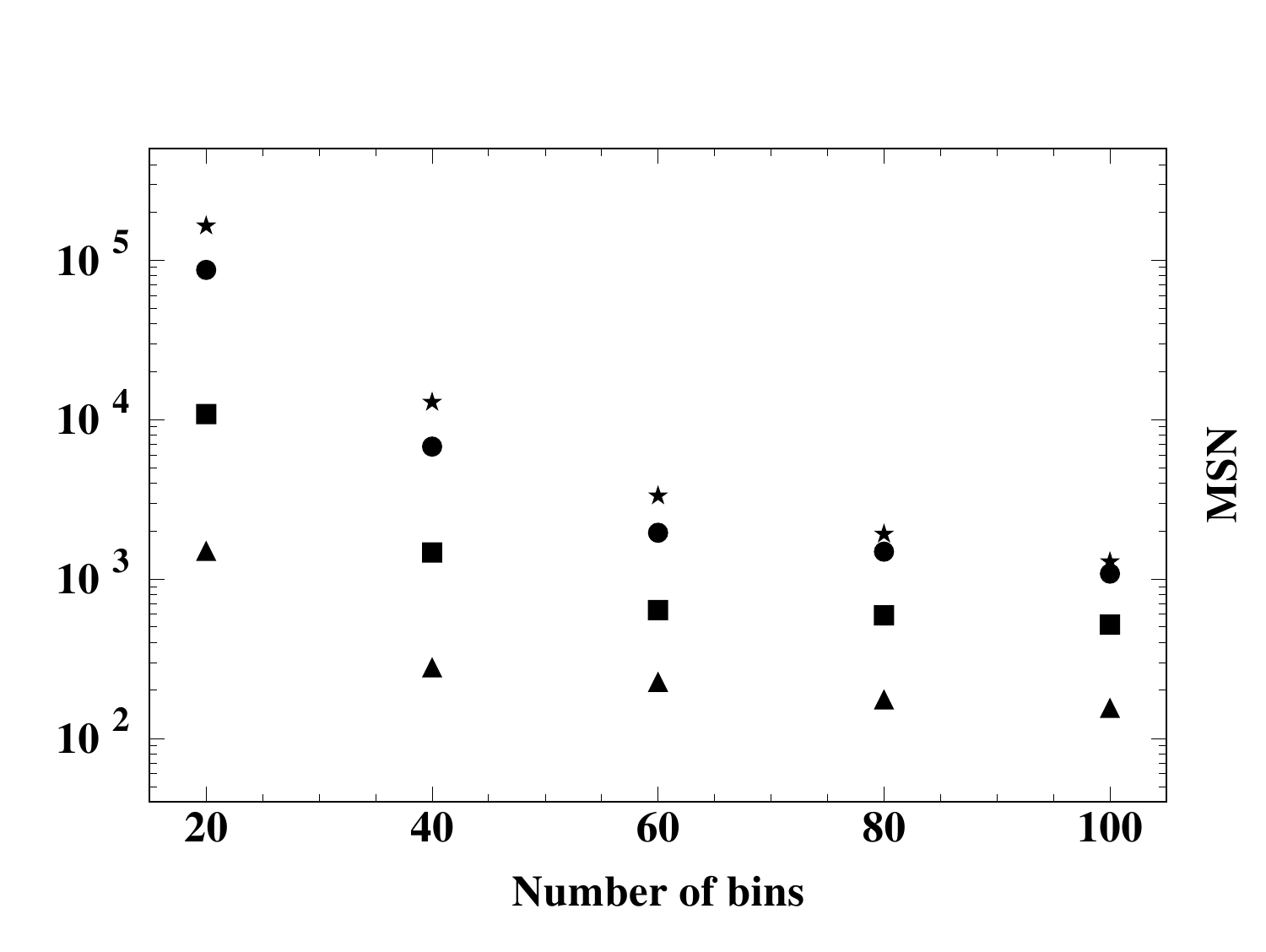}
	\end{subfigure}   
\vspace *{-0.2cm}      
	\caption{MSN for optimal value of MISE for case of NG method (left) and for RL method (right).}
	\label{fig:msn_eq}
\end{figure} 
Figure ~\ref{fig:msn_eq} shows that the MSN decreases as the number of bins in the measured distribution decreases to 100, at the optimal MISE, for both methods. The minimum MSN values are achieved with 15 bins in the unfolded distribution. In all cases, the MSN values obtained using the RL method are slightly lower than those obtained using the NG method.
\begin{figure}[H]
	\centering
 	\begin{subfigure}{0.495\linewidth}
		\includegraphics[width=\linewidth]{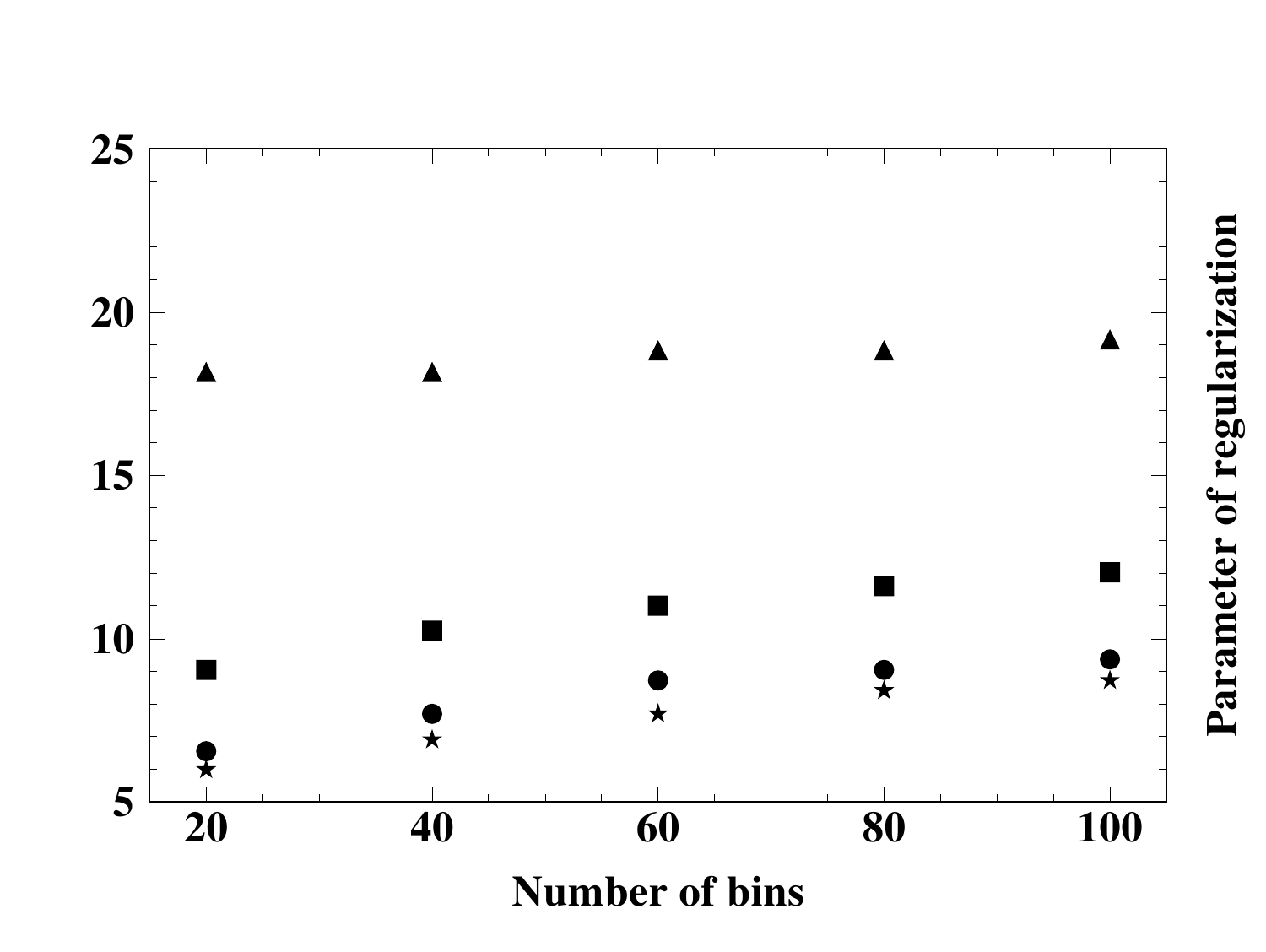}
	\end{subfigure}
   \hfill 
	\begin{subfigure}{0.495\linewidth}
		\includegraphics[width=\linewidth]{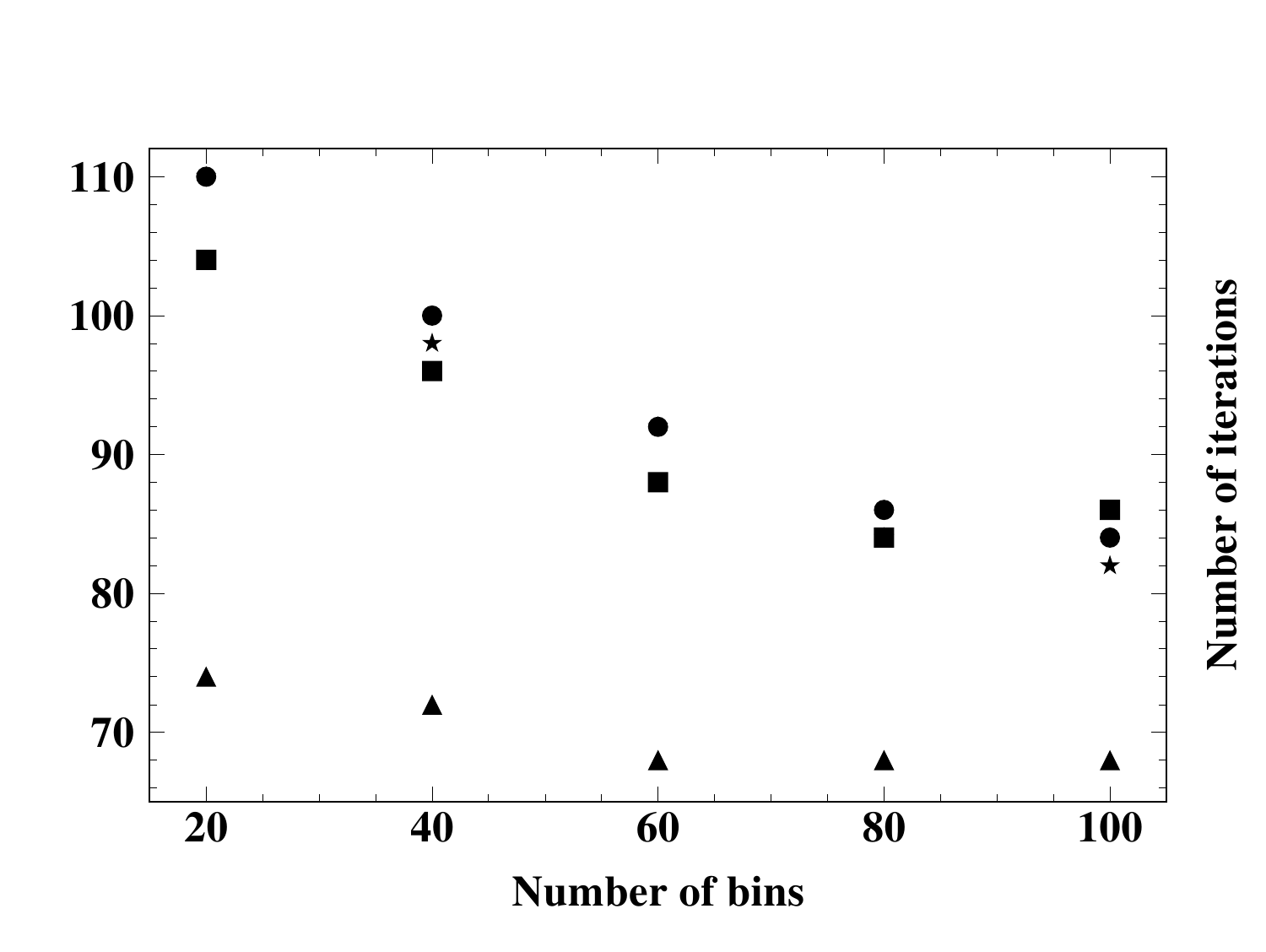}
	\end{subfigure}   
\vspace *{-0.2cm}      
	\caption{Regularization parameters for optimal value of MISE for case of NG method (left) and for RL method (right).}
	\label{fig:param_eq}
\end{figure} 
As shown in Figure~\ref{fig:param_eq}, the regularization parameter for the NG method increases with the number of bins in the measured distribution, while the number of iterations for the RL method decreases.

Based on a combination of parameters used to assess the quality of the results, the NG method produces better results than the RL method.
\newpage
To continue analysis the average unfolded distribution  were calculated for the case  80 bins in the measured distribution and 35 bins for unfolded distribution.  

\begin{figure}[H]
	\centering
 	\begin{subfigure}{0.495\linewidth}
		\includegraphics[width=\linewidth]{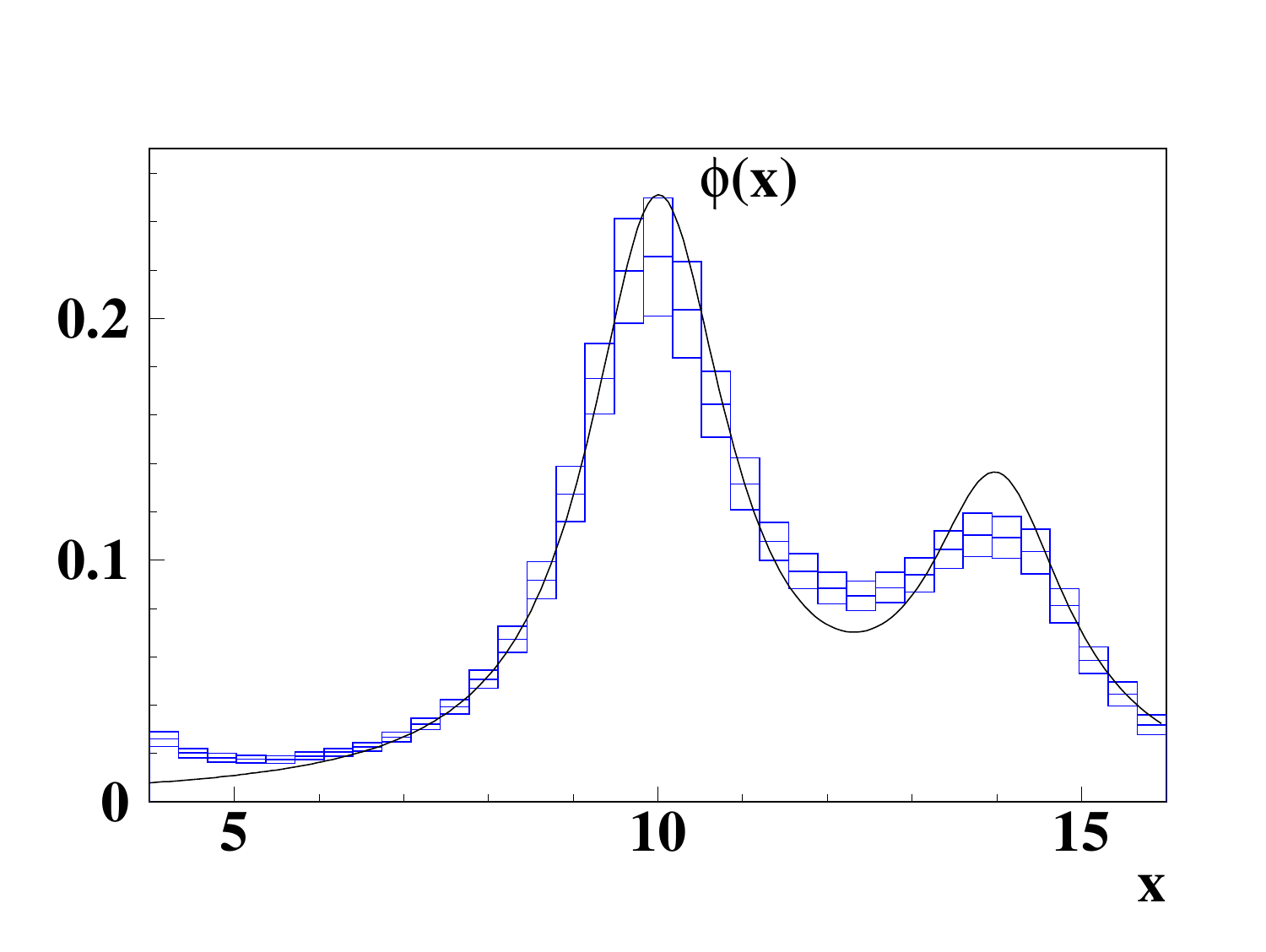}
	\end{subfigure}
   \hfill 
	\begin{subfigure}{0.495\linewidth}
		\includegraphics[width=\linewidth]{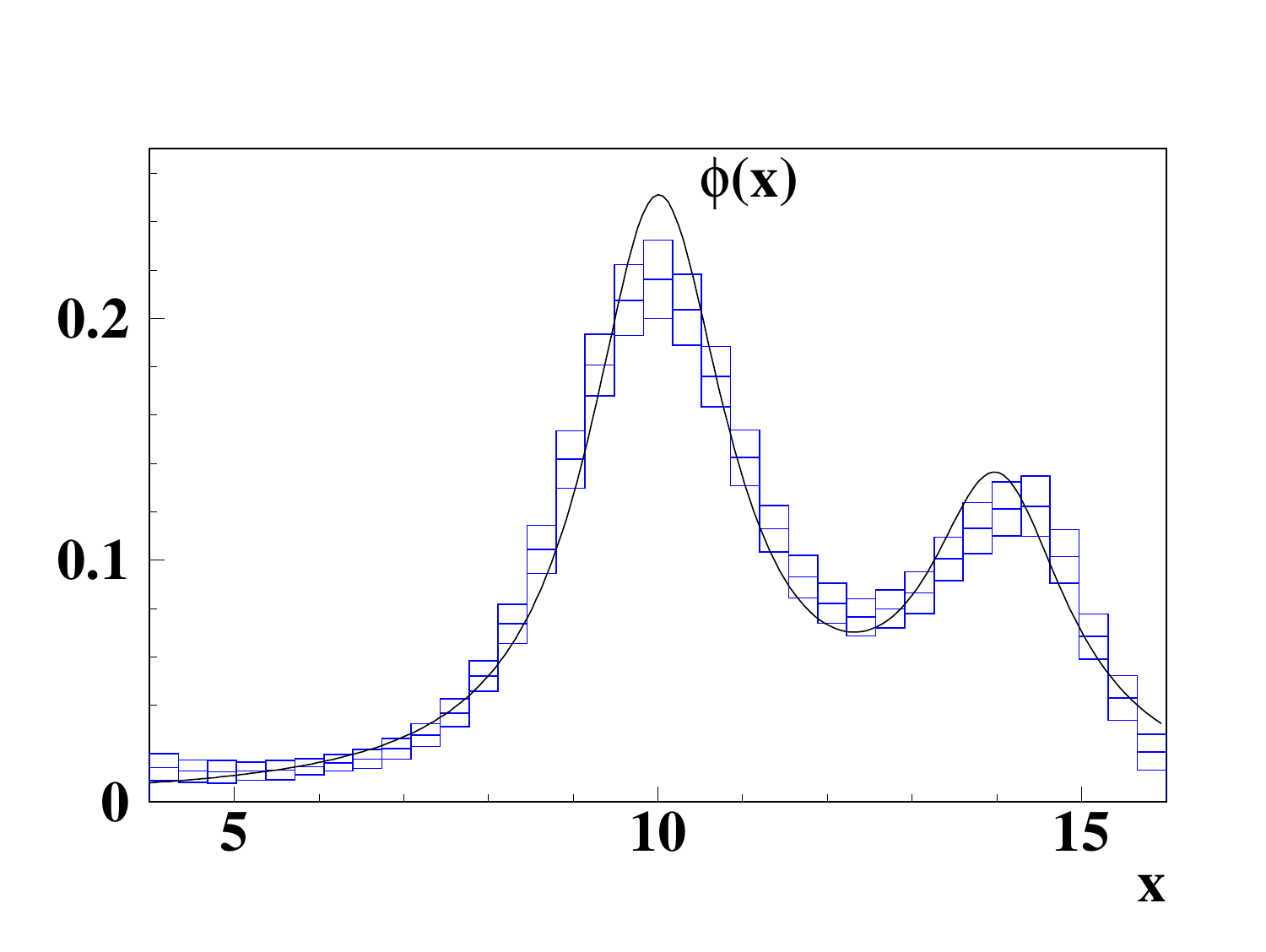}
	\end{subfigure}        
	\caption{Average unfolded PDF for NG method with the minimum MISE value (left) and  average unfolded PDF  for RL method with the minimum MISE value (right).}
	\label{fig:unfol_eq}
\end{figure}

\begin{figure}[H]
	\centering
 	\begin{subfigure}{0.495\linewidth}
		\includegraphics[width=\linewidth]{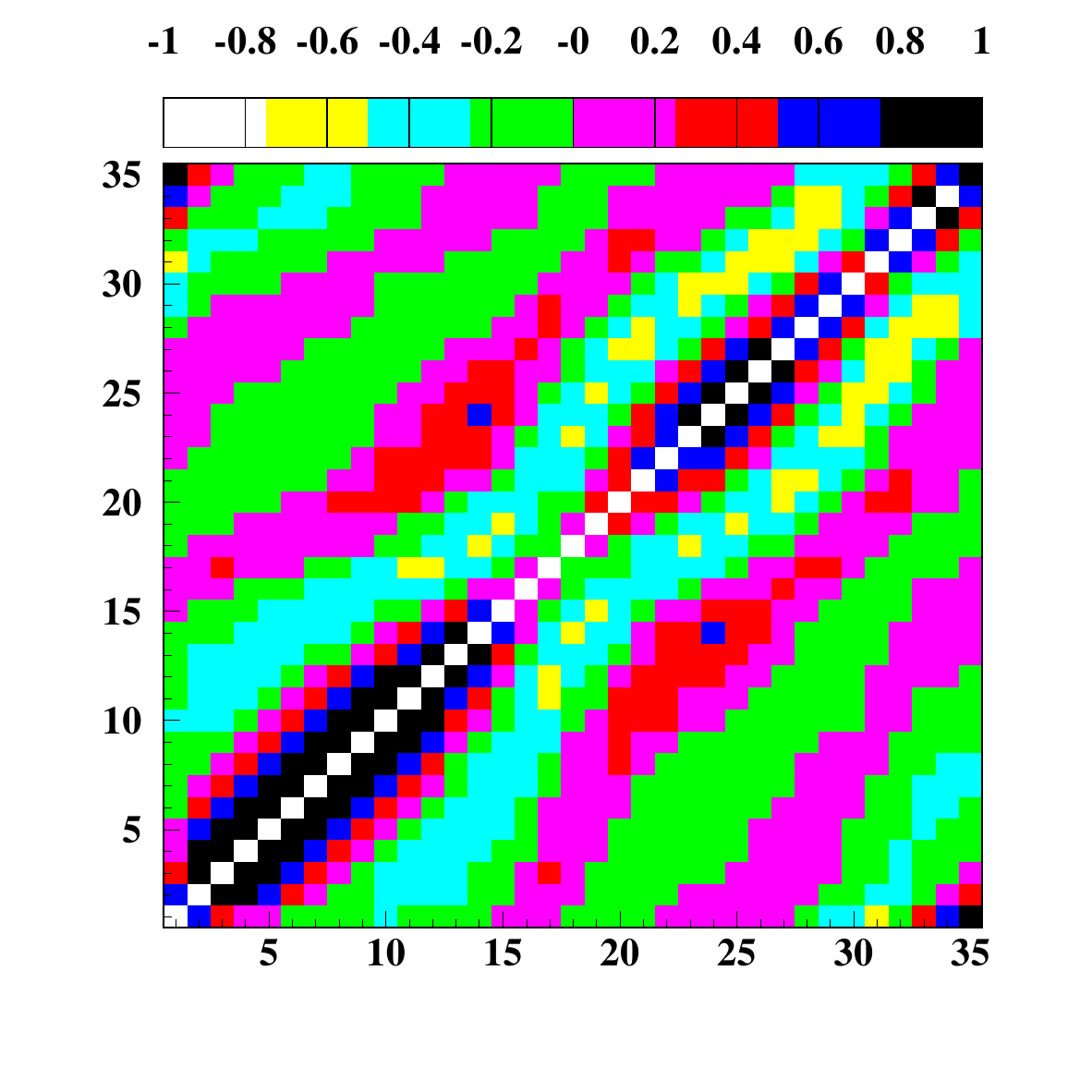}
	\end{subfigure}
   \hfill 
	\begin{subfigure}{0.495\linewidth}
		\includegraphics[width=\linewidth]{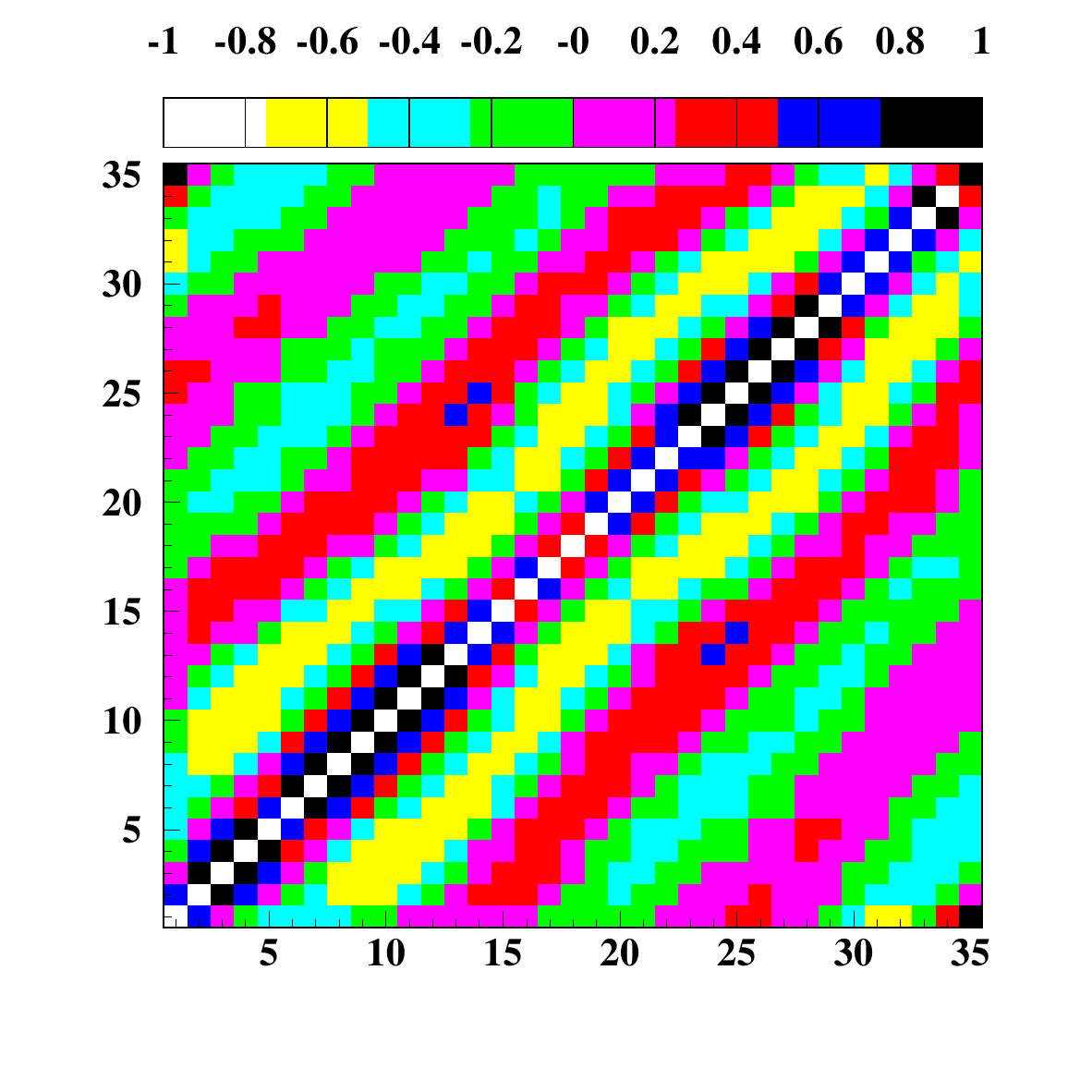}
	\end{subfigure}        
	\caption{ Correlation matrix for NG method  (left) and  correlation matrix for RL method (right)  }
	\label{fig:corr_eq}
\end{figure}

 The measured distribution with 80 bins  is shown in Figure~\ref{fig:true} and is used for comparing both methods.
The average unfolded distribution with 35 bins  is shown in  Figure~\ref{fig:unfol_eq}. 
The error bar represents the standard deviation of the estimate $\hat{\phi_i}/(a_{i+1}-a_{i})$, for bin $i$.
\vspace *{-0.5cm} 
The corresponding correlation matrices for both methods are also presented in Figure~\ref{fig:corr_eq}.
In addition, a table summarizing the optimal MISE, VISE and MSN values, together with the corresponding regularization parameters, is provided that again demonstrate prevalence of new data unfolding with MISE optimization under Richardson-Lucy method.
\begin{table}[H]
\begin{center}
\captionof{table}{ Characteristics of the optimal  MISE, VISE and  MCN for NG and RL algorithms  in case 35 equidistant bins for unfolded distribution and 80 bins for measured  one. }
\vspace {0.1 cm}
\begin{tabular}{ | c | c c c c  |  }
\hline
 Method &  MISE  & VISE&   MCN & Reg. par. \\
\hline 
  NG &  $2.843 \times 10^{-2}$  & $4.835 \times 10^{-4}$  & 5155.6 &9.0 \\
  RL &  $3.035 \times 10^{-2}$ & $7.394 \times 10^{-4}$  & 1480.8  &86\\
\hline
\end{tabular}
\end{center}
\end{table}
\section{Summary of section 6}
Figure~\ref{fig:mise_cl} and Figure~\ref{fig:mise_eq} show that the minimum MISE values are achieved with 35 bins in the unfolded distribution for both non-equidistant and equidistant binning, regardless of the unfolding method used.

For the optimal MISE values:

\begin{itemize}
\item {The VISE reaches its lowest value for the NG method using 35 non-equidistant bins. In all other cases, the minimum VISE value is achieved with 15 bins.}

\item {The MSN decreases as the number of bins in the measured distribution increases to 80, while its minimum value is achieved with 15 bins in all cases.}

\item {The regularization parameter for the NG method increases as the number of bins in the measured distribution increases, whereas the number of iterations for the RL method decreases.}
\end{itemize}

Finally, the NG method, when applied to k-means clustering-based binning with 35 bins, yields a substantially lower MISE, as shown in Figure~\ref{fig:mise_cl} (left). It also achieves lower VISE values and a relatively low MSN value, indicating the stability of the results.
\section{Conclusions}
Unfolding is considered a procedure for estimating an unknown probability density function. Different methods for assessing the quality of unfolding results are discussed. The main factors and parameters influencing the quality of the results are identified and formulated. 

Two methods based on likelihood-function optimization are considered: the Richardson–Lucy method, which uses the number of iterations as a regularization parameter, and a new data-unfolding method based on mean integrated squared error (MISE) optimization with entropy regularization.

A numerical example is presented to investigate the choice of binning as an important factor affecting the quality of unfolding results. The quality of the estimation is assessed in terms of the Mean Integrated Squared Error (MISE), the Variance of Integrated Squared Error (VISE)  and the Minimal Condition Number (MCN) of the correlation matrix of the estimated distribution.

A comparison is made between the best results obtained by varying the regularization parameters and the binning schemes. In both cases, the new method for estimating the response matrix is employed.

The numerical example demonstrates the substantial superiority of the data-unfolding method based on mean integrated squared error (MISE) optimization combined with k-means binning over the widely used Richardson–Lucy deconvolution method.

Presenting the unfolded distribution together with an appropriate assessment of the quality of the unfolding results significantly enhances the physical interpretation of experimental data.

\section {Acknowledgments}
The authors express their gratitude to the staff of the Icelandic Research E-Infrastructure for providing access to high-performance computing clusters, which were used to perform the computationally intensive calculations presented in this work.
\section *{References}
\bibliography{project}
\end{document}